\documentclass{aastex62}

\usepackage{amsmath,amsfonts,amssymb}
\usepackage{graphicx}
\usepackage{nicefrac}
\usepackage{mathtools}
\usepackage{gensymb}
\DeclareMathAlphabet{\mathbfsf}{\encodingdefault}{\sfdefault}{bx}{sl}

\newcommand{\vek}[1]{\boldsymbol{#1}}
\newcommand{\tens}[1]{\mathbfsf{#1}}

\newcommand{\pdq}[2]{\frac{\partial #1}{\partial #2}}

\graphicspath{{./}{figures/}}

\received{}
\revised{}
\accepted{\today}
\submitjournal{ApJ}

\shorttitle{COCONUT: Impact of the input magnetic map}
\shortauthors{Perri et al.}

\begin{document}

\title{COCONUT, a novel fast-converging MHD model for solar corona simulations: \\ II. Assessing the impact of the input magnetic map on space-weather forecasting at minimum of activity}

\correspondingauthor{Barbara Perri}
\email{barbara.perri@kuleuven.be}

\author[0000-0002-2137-2896]{Barbara Perri}
\affil{Centre for mathematical Plasma Astrophysics/Dept.\ of Mathematics, \\
KU Leuven, 3001 Leuven, Belgium}

\author[0000-0001-9438-9333]{B{\l}a{\.{z}}ej Ku{\'{z}}ma}
\affiliation{Centre for mathematical Plasma Astrophysics/Dept.\ of Mathematics, \\
KU Leuven, 3001 Leuven, Belgium}

\author[0000-0003-0874-2669]{Michaela Brchnelova}
\affiliation{Centre for mathematical Plasma Astrophysics/Dept.\ of Mathematics, \\
KU Leuven, 3001 Leuven, Belgium}

\author[0000-0002-1986-4496]{Tinatin Baratashvili}
\affiliation{Centre for mathematical Plasma Astrophysics/Dept.\ of Mathematics, \\
KU Leuven, 3001 Leuven, Belgium}

\author[0000-0002-9425-994X]{Fan Zhang}
\affiliation{Centre for mathematical Plasma Astrophysics/Dept.\ of Mathematics, \\
KU Leuven, 3001 Leuven, Belgium}

\author[0000-0003-3792-0452]{Peter Leitner}
\affiliation{Institute of Physics, University of Graz \\
Universit\"atsplatz~5, 8010 Graz, Austria}

\author[0000-0003-4017-215X]{Andrea Lani}
\affiliation{Centre for mathematical Plasma Astrophysics/Dept.\ of Mathematics, \\
KU Leuven, 3001 Leuven, Belgium}

\author[0000-0002-1743-0651]{Stefaan Poedts}
\affiliation{Centre for mathematical Plasma Astrophysics/Dept.\ of Mathematics, \\
KU Leuven, 3001 Leuven, Belgium}
\affiliation{Institute of Physics, University of Maria Curie-Sk{\l}odowska,\\
Pl.\ M.\ Curie-Sk{\l}odowskiej 5, 20-031 Lublin, Poland}

\begin{abstract}

% New abstract due to word limit
This paper is dedicated to the new implicit unstructured coronal code COCONUT, which aims at providing fast and accurate inputs for space weather forecast as an alternative to empirical models. We use all 20 available magnetic maps of the solar photosphere covering the date of the $2^{nd}$ of July 2019 which corresponds to a solar eclipse on Earth. We use the same standard pre-processing on all maps, then perform coronal MHD simulations with the same numerical and physical parameters. In the end, we quantify the performance for each map using three indicators from remote-sensing observations: white-light total solar eclipse images for the streamers' edges, EUV synoptic maps for coronal holes and white-light coronagraph images for the heliospheric current sheet. We discuss the performance for space weather forecasts and we show that the choice of the input magnetic map has a strong impact. We find performances between 24\% to 85\% for the streamers' edges, 24\% to 88\% for the coronal hole boundaries and a mean deviation between 4 to 12 degrees for the heliospheric current sheet position. We find that the HMI runs are globally performing better on all indicators, with the GONG-ADAPT being the second-best choice. HMI runs perform better for the streamers' edges, GONG-ADAPT for polar coronal holes, HMI synchronic for equatorial coronal holes and for the streamer belt. We especially showcase the importance of the filling of the poles. This demonstrates that the solar poles have to be taken into account even for ecliptic plane previsions.
\end{abstract}


\section{Introduction} \label{sec:intro}

% Space weather
With our societies increasingly relying on technology, we have now the critical need to anticipate major malfunctioning or even catastrophic events in order to protect civilians. Some of the most significant risks have been realized to be events coming from space \citep{Schrijver2015}. Highly energetic particles can be accelerated at the Sun or by magnetic structures in the interplanetary medium \citep{Reames_2013}, reaching energies that allow them to disrupt satellites, jeopardize astronauts' lives and interact with the Earth's atmosphere, leading to communication blackouts \citep{Bothmer2007}. These events are called Solar Energetic Particles events (SEPs); for more details, see review by \cite{Reames2021}. Magnetic storms are another type of events caused by coronal mass ejections (CMEs) interacting with Earth's magnetosphere \citep{Pulkkinen2007}, and resulting in currents in the Earth's crust that cause severe electrical damage to installations \citep{Pirjola2005}. Space weather has the mission to anticipate these disrupting events by simulating the chain of causality from the Sun to Earth and issue forecasts \citep{Temmer2021}. The key to reliable previsions is not only to be able to model accurately the transient phenomena, but to also describe precisely the interplanetary medium in which they propagate and with which they interact before reaching Earth \citep{Shen2022}. Although there are many effects that influence the transients' propagation \citep{Lavraud2014}, they can be linked back to two main physical ingredients. On the one hand, the magnetic field bathes the interplanetary medium, following a complex pattern influenced by the Parker spiral at large scales and fluctuations at small scales \citep{Owens2013}. Its long-term variations are linked to the 11-year cycle of solar activity generated inside the star by dynamo effect \citep{Brun2017}, while its short-term variations may be linked to the convection at the surface of the star \citep{Fargette2021}. On the other hand, the solar wind flows the interplanetary medium with continuous ejected plasma, and shapes large-scale structures with shock regions caused by the interaction between slow and fast wind (SIRs for Stream Interacting Regions) \citep{McComas2003, McComas2008}.

% Frameworks
It is only natural that an increasing number of countries are then developing frameworks for space weather forecasting: we can cite ENLIL and SWMF for the United States \citep{Odstrcil2003, Toth2012}, SUSANOO for Japan \citep{Shiota2014} and the VSWMC for Europe \citep{Poedts2020_vswmc}. All these frameworks are based on the same principle: since it is impossible to use one model to cover the diversity of scales between the Sun and Earth, the best approach is to couple models dedicated to a specific region and physics. For instance, the VSWMC framework uses photospheric measurements of the solar magnetic field as input, then semi-empirical (WSA) and magnetic (PFSS + SCS) extrapolations from 1 to $21.5\;R_\odot$, and then the heliospheric propagator EUHFORIA to compute physical quantities all the way from 0.1 AU to Earth and beyond (typical outer boundary condition is set at 2 AU) \citep{Pomoell2018}. The first steps of this chain of model, namely the magnetic map chosen as input and the coronal model used to compute the boundary conditions at 0.1 AU, are thus crucial as they determine the initialization of the rest of the models. They are also at the core of the two main physical ingredients that are going to disturb the transients' propagation: the magnetic maps are a direct measurement of the solar activity, and the solar corona is the siege of the acceleration of the solar wind \citep{Cranmer2019}.
% Revue of coronal models
To better model these sensitive effects, it is planned to use alternative magneto-frictional and MHD coronal models with more physics incorporated within, in order to replace and improve the semi-empirical and potential extrapolations up to 0.1 AU \citep{Poedts2020_euhforia}. Within the MHD models, there are other levels of complexity, such as the number of dimensions which are considered (1D vs. 3D) \citep{Pinto2017, Mikic2018}, or the level of sophistication to describe the coronal heating (polytropic vs. Alfvén waves) \citep{Perri2018, Reville2020}. There are even models going beyond the fluid approximation by taking into account the multi-species nature of the solar wind \citep{vanderHolst2014, Chhiber2021}. This approach has already proven successful for specific test cases \citep{Samara2021}. The dilemma is that, as we put more and more physics, what we gain in accuracy is lost in speed and robustness. As space weather forecasting requires all three qualities, we have developed a new coronal model to satisfy all these constraints. The COCONUT (COolfluid COroNal UnsTructured) coronal model uses the time-implicit methods from the COOLFluiD framework, which allows it to be up to 35 faster than typical explicit codes while achieving the same level of accuracy \citep{Perri2022}. It also has the advantage of using unstructured meshes instead of regular grids, which allow it to avoid degeneracy at the poles and thus provide more accuracy in this region. As more and more coronal models begin to be suited for space weather forecasts, another important effort for the community is to come up with metrics to evaluate the quality of the models and thus retain the best parameters for previsions \citep{Lionello2009, Wagner2022, Samara2022, Badman2022}.

% Impact of synoptic maps
This paper will focus in particular on the choice of the input magnetic map, as it is the driver of the entire numerical simulation. Many studies have tried to bridge the gap between various magnetic maps from different observatories, but no general consensus could be found behind these observations \citep{riley2014, virtanen2017}. This comes essentially from the lack of multi-vantage point, as for example no 360 degrees view of the Sun is available at all time since the breakdown of STEREO-B. New studies suggest that the choice of the input map and its pre-processing would change significantly the description of the coronal structure \citep{Yeates2018}, and thus of the SIRs and CME propagation \citep{Riley2021, Samara2021}. For this reason, more and more studies focus on trying to assess the impact of the choice of the input map on the resulting coronal structure \citep{Petrie2013, Wallace2019, Caplan2021, Li2021}. However, most of these studies rely on PFSS extrapolations to describe the coronal magnetic field, while MHD would be more physical, especially further away from the star \citep{Reville2015b}. MHD studies have started to be conducted, but so far mostly for few codes, which are the MAS and AWSoM codes \citep{Linker2017, meng2022}.
For all magnetic maps, the greatest uncertainty lies in the solar poles, as the viewpoint from Earth and satellites in the ecliptic plane does not allow for precise global measurement. Only local observations by Hinode or soon Solar Orbiter allows us to retrieve high-resolution information from the solar poles \citep{Tsuneta2008}. There are however indirect techniques that can be used such as microwave imaging observations \citep{Gopalswamy2012} or Zeeman effect \citep{Ito2010}. This is problematic for global coronal models, as it leads to huge uncertainties on the open solar flux \citep{Riley2019} and therefore underestimation of the magnetic field at Earth \citep{Owens2008, Jian2015}. The solar poles have been known to influence greatly the dynamics of the corona, by affecting the IMF field strength, the HCS excursions, and the wind speed through the polar coronal holes \citep{Petrie2015}. However, the impact of the solar poles modeling in space weather forecasts is still not properly quantified. It is made even more difficult by the fact that most models do not include the solar poles in the heliospheric part \citep{Pomoell2018}, and sometimes even in the coronal part \citep{Pinto2017}, thus implicitly assuming the influence of the poles can be neglected.
Our goal is to test these assumptions, first for a well-documented case of minimum of activity of the $2_{nd}$ of July 2019. The choice of the minimum of activity allows us to focus on the influence of the poles rather than the active regions, which is also made possible by our unstructured mesh approach allowing for fully including the poles within the computational domain. The choice of the date allows us to have precise pictures of the solar corona thanks to a total solar eclipse as seen from Earth.
  
% Outline of the paper
This paper is organized as follows. In section \ref{sec:mag_maps}, we give an overview of the magnetic maps which are used as input of our simulations (all 20 maps publicly available for the $2^{nd}$ of July 2019 total solar eclipse), explaining in particular their differences in spectral line selection, resolution and pole-filling techniques. In section \ref{sec:cf}, we then present our numerical model COCONUT which uses these magnetic maps in order to simulate the solar wind in the corona up to 0.1 AU. We describe the physical as well as the numerical parameters which are used to constrain the simulations. We also discuss the pre-processing of the maps for quantifying the difference in initialization of the simulations. In section \ref{sec:comp_min}, we analyze the results of the 20 corresponding simulations which have been performed. We use 3 different observational data available for this date to validate the results: we compare magnetic field lines to white-light images (section \ref{subsec:min_wl}), open and closed magnetic field line distribution to coronal hole detection in EUV (section \ref{subsec:min_ch}) and the position of the Heliospheric Current Sheet (HCS) to the Streamer Belt (SB) white-light reconstruction (section \ref{subsec:min_hcs}). In section \ref{sec:discussion}, we discuss the implications for space weather forecasting. We begin by comparing the resulting magnetic field configuration at 0.1 AU with the typical WSA + PFSS + SCS model used currently for coupling with EUHFORIA (section \ref{subsec:min_space_weather_forecast}). We then assemble all our results into a scoreboard for this event, determining which magnetic map allows our model to fit the observational data the best (section \ref{subsec:map_scores}). We focus especially on the pole-filling techniques and their implication for forecasts (section \ref{subsec:poles}). Finally, in section \ref{sec:conclusion} we sum up the conclusions of our study and present the perspectives for future work. 

\section{Description of the magnetic maps} 
\label{sec:mag_maps}

% General intro
Our simulations are data-driven in the sense that the inner boundary condition for the radial magnetic field $B_r$ is imposed based on a synoptic map derived from solar observations of the photospheric magnetic field. There are also models which are fully data-driven because they use the three components of vector magnetograms as an inner boundary condition, along with velocity components $V_\theta$ and $V_\varphi$. The number of Dirichlet conditions is then determined by the directions of the characteristic waves going in and out of the photosphere \citep{Wu2006, Yalim2017, Singh2018}. Such methods are more difficult to implement within our unstructured grid and implicit solver, so this remains outside the scope of this study and will be considered for future extensions of the code. For the selected date ($2^{nd}$ of July 2019), we used all publicly available processed synoptic maps from 4 different providers: WSO (Wilcox Solar Observatory), GONG (Global Oscillation Network Group), HMI (Helioseismic and Magnetic Imager) and GONG-ADAPT (Air Force Data Assimilative Photospheric Flux transport). Links to their corresponding source in order to download them are shown in the acknowledgments section. A summary of their main properties can be found in table \ref{tab:maps}. In this section we will explain the differences between these different maps, focusing on the observation techniques, the assembly methods and the pole-filling methods.

\begin{table}[!t]
    \centering
    \begin{tabular}{|c||c|c|c|c|c|c|c|c|}
        \hline
        Provider & Spectral line & Type & Resolution & Units & Y-axis & Pole filling & Time span & CRs \\ \hline \hline
        WSO & Fe 525 nm & LOS & 73x30 & $\mu T$ & $\rm{sin}\theta$ or $\theta$ & None & 1976.3 - & 1642 - \\ 
        GONG & Ni 676.8 nm & Pseudo-Radial & 360x180 & $G$ & $\rm{sin}\theta$ & Cubic-polynomial fit & 2006.7 - & 2047 - \\ 
        HMI & Fe 617.3 nm & Pseudo-Radial & 3600x1440 & $G$ & $\rm{sin}\theta$ & None & 2010.4 - & 2096 - \\ 
        GONG-ADAPT & Ni 676.8 nm & Pseudo-Radial & 360x180 & $G$ & $\theta$ & Flux-transport models & 2007.002 - & 2052 - \\ 
        \hline
    \end{tabular}
    \caption{Properties of synoptic magnetic maps which have been used in this study. For each provider, we specify the observed spectral line, the type of magnetic field, the resolution of the map, the units of the magnetic field, the type of $y$ axis which has been used, the pole-filling technique, the available time span and the corresponding Carrington Rotations (CRs). For the source of the magnetic maps, please check the acknowledgments section.}
    \label{tab:maps}
\end{table}

% Description of the various techniques for measurement
All the maps were obtained through magnetographs, although the latter use various techniques in different contexts. 
% Spectral line
A first difference is the observed spectral line, as seen in column 2 of table \ref{tab:maps}. At WSO, a Babcock solar magnetograph records the Zeeman polarization in the wings of an absorption line of iron at 5250 \AA \citep{Ulrich1992}. It is the longest homogeneous series of observations with the same instrumentation which has been used since 1976. GONG uses interferometric techniques in order to measure the opposite states of polarization of the Ni I 6768 \AA line, which is based on 6 stations around the world since 2006. HMI is an instrument onboard the SDO satellite (Solar Dynamics Observatory) launched in 2010, which observes the full solar disk in the Fe I absorption line at 6173 \AA. It was calibrated using the instrument MDI (Michelson Doppler Imager) onboard SOHO (Solar and Heliospheric Observatory). It can also record 3D vector magnetograms. Finally, the GONG-ADAPT maps are based on the GONG observations, so relying as well on the Ni I 6768 \AA line. These differences in spectral line technically mean that the maps are not representing the magnetic field at the same height, which can result in slightly different structures.
% Explanation of type
The third column refers to the fact that all observatories measure line-of-sight (LOS) component of the magnetic field. However some of them convert this value into a pseudo-radial field under the assumption that the total field is radial.
% Resolution
Column 4 shows another important difference between the maps which is their resolution. WSO is the lowest-resolution device with only a 3-arcmin aperture size, which results in maps of 73 pixels in longitude and 30 pixels in latitude. GONG (and consequently GONG-ADAPT) provides map products with 360 pixels in longitude and 180 pixels in latitude. Finally, HMI has the best resolution thanks to the fact that it is in space, with a 1-arc-second resolution, and provides high-resolution maps with 3600 pixels in longitude and 1440 pixels in latitude.
% Units and y-axis
We also note in column 5 that the units are mostly in Gauss, except for the WSO maps which are in micro-Teslas. Column 6 shows another important geometric parameter which is the type of $y$ axis used. "$\theta$" means that the pixels are in equal steps of latitude, which is the case for GONG-ADAPT between -90 and 90 degrees, and a possible option for WSO between -70 and 70 degrees. "$\rm{sin}\theta$" means that the pixels are in equal steps of sine latitude (to account for the fact that the poles are difficult to measure from the point of view of the ecliptic plane), which is the case for GONG and HMI between -1 and 1, and an option for WSO between -14.5/15 and 14.5/15.
% Processing
We should also note that over the years, various processings have been applied to the data or have been highly recommended. In this study, we took the maps as they were, and chose to not apply any correction. WSO for example had several periods with sensitivity issues, some of them having been recalibrated (between November 200 and July 2002, and between $16^{th}$ December 2016 and $18^{th}$ May 2017). There is also a general problem of saturation described in \cite{svalgaard1978} and updated in \cite{svalgaard2006good}. Please note that the difference between GONG and GONG-ADAPT is also mostly some post-processing, as we will explain in the next paragraph. This modification history is not always made public, and thus can produce differences based on the date at which the data have been downloaded and processed.
% References
For more details about the instruments, the reader can also refer to the reviews of \cite{riley2014} and \cite{virtanen2017}.

\begin{figure}
    \centering
    \includegraphics[width=\textwidth]{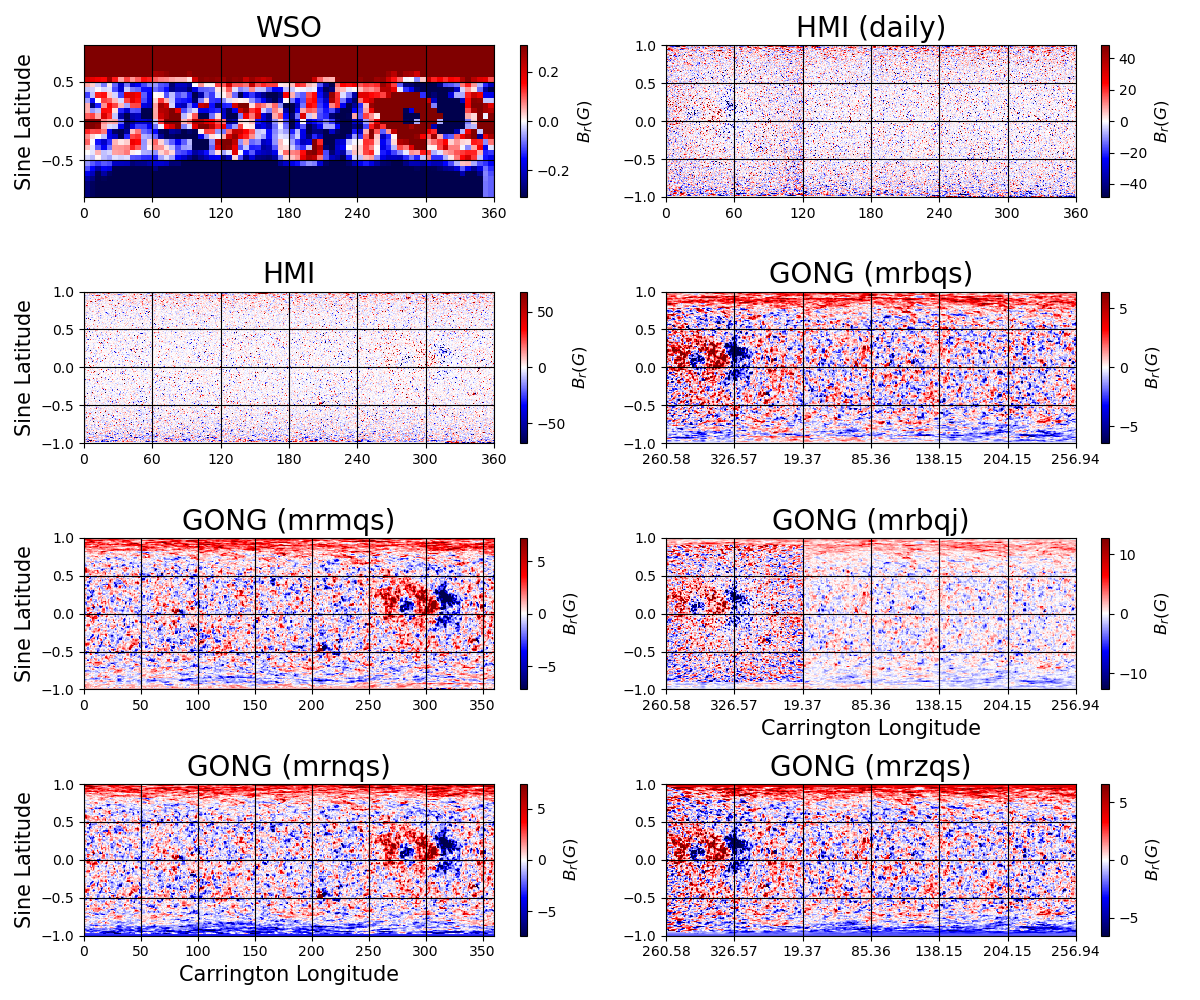}
    \caption{Comparison of synoptic maps for the $2^{nd}$ of July 2019 (CR2219). From top to bottom, and then left to right: WSO, HMI, GONG (mrmqs), GONG (mrnqs), HMI (synchronic), GONG (mrbqs), GONG (mrbqj), and GONG (mrzqs). The first column shows Carrington-frame synoptic maps, while the second column shows maps with longitude converted to the Carrington longitude for CR2219. All data are in their original resolution and axis (longitude - sine-latitude). The ranges of the color bars have been set to plus and minus of the maximum of the field divided by 10, in order to have positive polarities in red and negative polarities in blue, as well as a good balance between small and large-scale structures.}
    \label{fig:maps_other}
\end{figure}

\begin{figure}
    \centering
    \includegraphics[width=\textwidth]{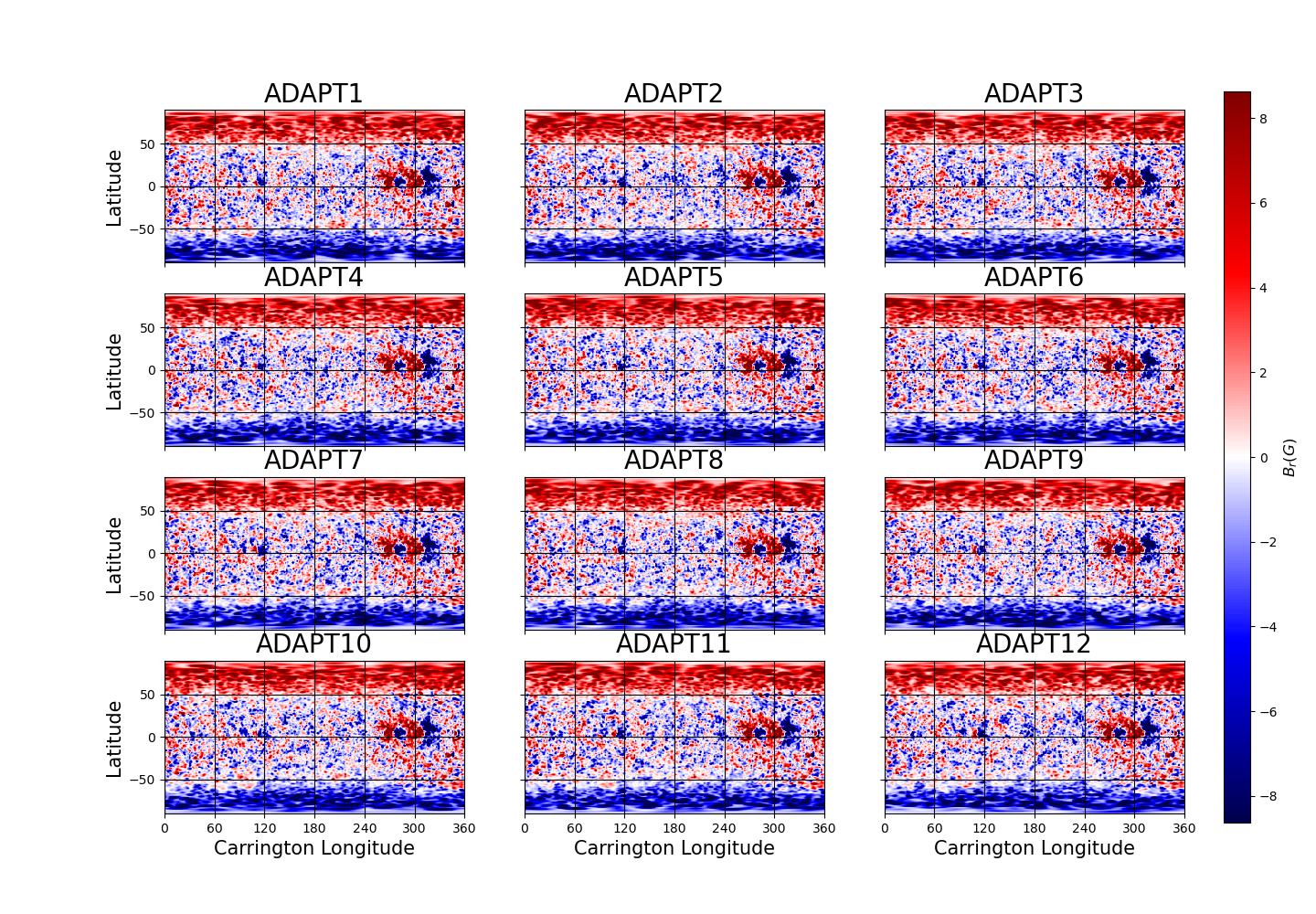}
    \caption{Comparison of the 12 GONG-ADAPT realizations for the $2^{nd}$ of July 2019 (CR2219). All data are in their original resolution and axis (longitude - latitude). The ranges of the color bars have been set to plus and minus of the maximum of the field divided by 10, in order to have positive polarities in red and negative polarities in blue, as well as a good balance between small and large-scale structures.}
    \label{fig:maps_adapt}
\end{figure}

% Synoptic vs synchronic
Another important difference to discuss is the way the synoptic maps are assembled, and the very definition of a synoptic map in the first place. A synoptic map means that the full surface of the Sun is covered in 360 degrees. However, it does not guarantee that all data which were used to create this full view were taken at the same time (this would be called a synchronic map). In reality, most of the maps are assembled using data at different dates, thus producing diachronic maps. For the WSO map, the full-disk images of the Sun are remapped over a month into Carrington longitudes, which means that there is a 27-day difference on average between data at 0 and 360 degrees on the map. The HMI map follows the same idea, except that the better resolution allows to average 20 magnetograms for each Carrington longitude. More precisely, individual pseudo-radial magnetograms are remapped on a very high-resolution Carrington coordinate grid. For each Carrington longitude, the 20 magnetograms closest in time to the central meridian passage (CMP) (within 2 degrees) for that longitude are selected and averaged. The result is that the effective temporal width of the HMI synoptic map is about three hours. The choice of a constant number of contributing magnetograms allows to minimize the variation of the noise over the entire map. A two-dimensional Gaussian function (whose width is 3 pixels) is then applied to high-resolution remapped data to reduce the spatial resolution before generating the high-resolution synoptic maps.\footnote{\url{http://jsoc.stanford.edu/HMI/LOS_Synoptic_charts.html}}
The HMI daily update synchronic frames provide a more up-to-date version of the synoptic map with the first 120 degrees being replaced by the daily full-disk observation at the corresponding date from the twenty 720s-magnetograms obtained between 10 and 14 UT, which helps reducing the time gap between data and allows to take into account fast-evolving structures. The origin of the frame is adjusted so that the newest data will appear on the left of the 360 degree map. We will refer to this frame as the synchronic frame through the rest of this article. It does not mean that the full map is synchronic, but it is chosen so that the central meridian of the given date is always at 60 degrees in longitude from the left-leading edge.
% Map specificities
Within this set of maps, we would like to take some time to describe more precisely some subsets of maps.
% Different GONG
Within the GONG products, there are 5 different synoptic maps available. 2 of them are integral magnetogram synoptic maps, and follow the same idea as described before: the mrmqs and mrnqs maps are built using data from the full Carrington rotation. To derive a map of the full-sun magnetic field, fully calibrated one-minute full-disk photospheric magnetograms from GONG's six sites are used. The first step is that the one-minute images from the GONG network are merged to give continuous minute-by-minute coverage of the field. Then the merged images are remapped into longitude (measured from the central meridian) and sine latitude. Next, these remapped images are shifted to the Carrington frame and merged with a weighted sum to form a full-surface picture of the solar magnetic field. Weighting factors take the form of a cosine to the power 4 of the longitude to ensure that measurements taken at a particular Carrington time contribute most to that Carrington longitude in the final synoptic map.\footnote{\url{https://gong.nso.edu/data/dmac_magmap/}} The 3 others are synchronic frames magnetogram synoptic maps. This is especially visible when we plot all the maps in figure \ref{fig:maps_other}. The mrbqj product called the Janus maps are similar to the HMI synchronic frame maps: the left 60 degrees in longitude between -60 and 60 degrees in latitude are updated using classic synoptic information, thus resulting in a composite magnetogram. However, in the case of the mrbqs and mrzqs products, this means that the 60 degrees to the left of the map have not crossed the central meridian, and are thus not updated for the current Carrington rotation. Then, there is another distinction made between the zero-point corrected products (mrzqs, mrnqs) and the standard products (mrbqs, mrbqj, mrmqs): these maps have corrections at the poles to have a better estimate of the global magnetic flux. This is visible in figure \ref{fig:maps_other} where we see the southern pole negative polarity being enhanced for GONG mrzqs and GONG mrnqs.
% Different ADAPT
Within the GONG-ADAPT map, there are actually 12 realizations produced. The differences rely on the various models used to try to approximate a synchronic map \citep{Hickmann2015}: here, GONG full-disk magnetograms are processed using forward modeling to account for differential rotation, meridional circulation and supergranulation. Combined with data assimilation, this leads to a model ensemble of 12 realizations at the time of observation. All these different realizations are plotted on figure \ref{fig:maps_adapt} for the $2^{nd}$ of July 2019 in order to show the differences for a minimum of activity.

\begin{table}[!t]
    \centering
    \begin{tabular}{|c||c|c|c|c|}
        \hline
        Name & Full name & Frame & Zero-point correction & Updated data \\ \hline
        mrmqs & Integral Carrington Rotation Magnetogram Synoptic Map & Carrington & no & no \\ \hline
        mrnqs & Integral synoptic map & Carrington & yes & no \\ \hline
        mrbqs & Standard QuickReduce Magnetogram Synoptic Map & Synchronic & no & no \\ \hline
        mrbqj & Janus QuickReduce Magnetogram Synoptic Map & Synchronic & no & yes \\ \hline
        mrzqs & Synoptic map & Synchronic & yes & no \\ 
        \hline
    \end{tabular}
    \caption{Summary of the properties of the GONG products. For each product, we explain the full name of the product and the associated frame. We also specify whether the zero-point correction is applied, and whether updated data are included.}
    \label{tab:gong_maps}
\end{table}

% GONG discussion
To make it easier for the reader, we have summarized the main properties of the various GONG products in Table \ref{tab:gong_maps}. Not all of these products were necessarily designed to be used as inputs for coronal modeling and space weather previsions. The recommended products are the zero-point corrected ones (mrzqs and mrnqs), but for practical reasons, it turns out that some facilities still use the non-corrected synchronic products (mrbqs) \citep{Poedts2020_vswmc}, which makes them still relevant to study. The Janus maps were designed to reproduce more closely sudden changes of magnetic flux in the solar disk facing Earth. This makes them more precise but also possibly more unstable because noisier. Finally, the integral maps in Carrington frame were not necessarily designed as an operational product, but they are closer to the HMI map, and we found it interesting to adopt an unbiased approach and test all of these maps for our model.

% Pole-filling
Finally, the maps may use different techniques to fill the solar poles. The solar poles are currently not clearly visible with an extended range of latitudes by any magnetograph because all of them are located in the ecliptic plane, perpendicular to the poles. This will change with Solar Orbiter, which is scheduled to go 30 degrees out of the ecliptic plane around 2025, in order to provide more detailed global pictures of the solar poles with an extended range of accessible latitudes. In the meantime, magnetic maps need to use extrapolation techniques if they want to improve the description of the poles. In the set that we are studying, we can see in table \ref{tab:maps} that the HMI map has no correction for the poles. The WSO map neither, but since it does not provide data between $-70^\circ$ and $-90^\circ$, and $70^\circ$ and $90^\circ$, we perform a linear extrapolation to fill these gaps. This means that the WSO map is going to have the least accurate information about the solar poles due to instrument limitations, since all data above 55 degrees of latitude comes from only one 3-arcmin pixel. The GONG map performs a cubic-polynomial fit. Finally, the GONG-ADAPT has the most sophisticated model, which takes into account flux-transport to increase the concentration of the magnetic field at the poles because of the modeled meridional circulation.

\section{Description of COCONUT code} 
\label{sec:cf}

% Basic description
COCONUT stands for COolfluid COroNa UnsTructured, and is a 3D MHD coronal model based on a fully implicit solver for Finite Volume Methods (FVM) on unstructured grids. The solver is part of the COOLFluiD framework (Computational Object-Oriented Libraries for Fluid Dynamics) \citep{Lani2005, Lani2006, Kimpe2005, Lani2013}, designed for scientific heterogeneous high-performance computing of multi-physics applications, including astrophysical plasmas \citep{LaniGPU, Laguna2016, Maneva2017, Laguna2018, Asensio2019}. We refer the reader to \cite{Perri2022} for the complete description of the COCONUT code. We will focus here on its main physical and numerical features.

\subsection{Equations and physical parameters}
\label{subsec:cf_physics}

% Equations
We solve the ideal MHD equations in conservation form in Cartesian coordinates (more details are given in \citet{Yalim, LaniGPU}):

\begin{equation}
\frac{\partial}{\partial t}\left(\begin{array}{c}
\rho \\
\rho \vek{v} \\
\vek{B} \\
E \\
\rho \\
\phi 
\end{array}\right)+\vek{\nabla} \cdot\left(\begin{array}{c}
\rho \vek{v} \\
\rho \vek{v} \vek{v}+\tens I\left(p+\frac{1}{2}|\vek{B}|^{2}\right)-\vek{B} \vek{B} \\
\vek{v} \vek{B}-\vek{B} \vek{v}+\underline{\tens I \phi} \\
\left(E+p+\frac{1}{2}|\vek{B}|^{2}\right) \vek{v}-\vek{B}(\vek{v} \cdot \vek{B}) \\
V_{r e f}^{2} \vek{B}
\end{array}\right)=\left(\begin{array}{c}
0 \\
\rho \vek{g}\\
0\\
0 \\
\rho \vek{g} \cdot \vek{v} \\
0
\end{array}\right),
\end{equation}
in which ${E}$ is the total energy, $\vek{B}$ is the magnetic field, $\vek{v}$ the velocity, $\vek{g}$ the gravitational acceleration, $\rho$ the density, and $p$ is the thermal gas pressure. The gravitational acceleration is given by $\vek{g}(r) = -(G M_\odot/r^2)\, \hat{\vek{e}}_r$ and the identity dyadic $ \tens I = \hat{\vek{e}}_x \otimes \hat{\vek{e}}_x + \hat{\vek{e}}_y \otimes \hat{\vek{e}}_y + \hat{\vek{e}}_z \otimes \hat{\vek{e}}_z$.

% Normalization
Since the ideal MHD equations are scale independent, they are implemented in COOLFluiD in dimensionless form. The following basis set $\{\ell_0,\rho_0,B_0\}$ of code units $Q_0$ is used to adimensionalize any physical quantity $Q$ as $\tilde Q = Q/Q_0$: the unit length $\ell_0 = R_\odot =6.95\times10^{10}\,\rm{cm}$, unit mass density $\rho_0 = \rho_\odot=1.67\times10^{-16}\,\rm{g\,cm^{-3}}$, and $B_0 = 2.2\ \mathrm{G}$, a typical value for the background solar dipole field all represent solar surface values. All other code units are composed of combinations of the three base units, such as unit pressure $P_0 = \rho_0 V_0^2$ and gravitational acceleration $g_0 = V_0^2/\ell_0$ with $V_0 = B_0/\sqrt{\mu_0 \rho_0}$.

% Input parameters
We use typical solar surface values for the mass density $\rho_\odot = 1.67 \times 10^{-16}\ \mathrm{g/cm^3}$ and $T_\odot = 1.9 \times 10^6\ \mathrm{K}$ for fixed-value Dirichlet conditions of density and pressure. The pressure at the inner boundary follows from the solar surface temperature by application of the ideal gas law: $P_\odot = 4.15 \times 10^{-2} \, \mathrm{dyn/cm^2}$.

\subsection{Numerical methods and boundary conditions}
\label{subsec:cf_numerics}

% Numerical methods
The state variables are evolved in time using a one-point and three-point implicit Backward Euler scheme for steady and unsteady cases \citep{Yalim}, respectively, solving the resulting linear system with the Generalized Minimal RESidual (GMRES) method  \citep{Saad1986} which is implemented within the PETSc library \citep{petsc-web-page, petsc-user-ref, petsc-efficient}.

% Divergence cleaning
In order to ensure the divergence constraint $\nabla \cdot \vek B = 0$, we use the Artificial Compressibility Analogy \citep{chorin1997}, which is very similar to the Hyperbolic Divergence Cleaning (HDC) method originally developed by \cite{Dedner2002} and has been shown to perform well with our implicit solver \citep{Yalim}:
\begin{equation} \label{eqn:hdc}
    \pdq \phi t + V_{ref}^2 \nabla \cdot \vek B = 0
\end{equation}
which couples the zero-divergence constraint to the induction equation, ensuring that the whole system remains purely hyperbolic. $c_h$ denotes the propagation speed of the numerical divergence error, set to $1.0$.

% Boundary conditions
% Inner boundary conditions
% Hydro variables
The inner velocity is set to 0 at the inner boundary by following the prescription: $V_{{x,y,z}G} = - V_{{x,y,z}I}$. This condition allows us to suppress the currents at the solar surface in order to produce a better perfect conductor boundary condition (see \cite{Perri2022} and \cite{Brchnelova2022b} for more details). 
%Rotation is set in the ghost cells to the standard solar rotation rate of 27 days.

% Magnetic field
In order to be able to pass an initial condition for the magnetic field distribution to the MHD solver, we compute a potential field approximation based on a particular magnetic map as inner (i.e. at the solar surface) boundary condition. From the input synoptic map, we derive a Dirichlet condition based on the radial magnetic field: $B_{r\mathrm{G}} = 2 B_{r\mathrm{PF}}\Big|_{\partial \varOmega_\mathrm{i}} - B_{r\mathrm{I}}$. Here and in the following, index~``G'' is supposed to indicate a value evaluated at a particular ghost cell center, while index ``I'' refers to the corresponding inner cell, adjacent to the ghost cell. The field value at the ghost cell center is assigned such that the exact boundary value at the cell face bordering ghost- and inner state symmetrically, e.g.\ $B_{r\mathrm{PF}}|_{\varOmega_\mathrm{i}}$ is the arithmetic mean of the quantity in question as evaluated on the ghost- and inner state cell centers. $\partial \varOmega_\mathrm{i} = \{(r,\vartheta,\varphi)|r=R_\odot\}$ denotes the solar surface boundary and $\partial \varOmega_\mathrm{o}$ the outer spherical shell at $r=21.5\;R_\odot$. Because the other components of the magnetic field are not derived from data, we use simple zero gradient conditions across the inner boundary ($\partial B_\theta/\partial r = \partial B_\varphi/\partial r = 0$).

% Outer boundary conditions
Due to the solar wind being supersonic at $r = 20.0\;R_\mathrm{S}$, we can extrapolate the spherical field components $r^2 B_r$, $B_\vartheta$, $B_\varphi$, as well as $\rho$, $V_r$, $V_\vartheta$, $V_\varphi$ and $P$ from the outermost cell centers to the ghost cells with a zero gradient. We extrapolate $r^2 B_r$ instead of $B_r$ to comply with the divergence-free constraint for the magnetic field (see \cite{Perri2018} for more details).

% Mesh
The mesh which has been used for all simulations is a spherical shell domain defined by: $\varOmega = \{(r,\vartheta,\varphi)|R_\odot < r < 21.5\;R_\odot\}$, where the inner and outer boundary conditions are applied on $r = R_\odot$ and $r = 21.5R_\odot$ respectively. 
The surface mesh of a level-6 subdivided geodesic polyhedron (consisting of triangular elements) was generated to represent the inner boundary and then extended radially outwards in layers until the outer boundary was reached, resulting in a 3-D domain consisting of prismatic elements. The default mesh used a 6th-level subdivision of the geodesic polyhedron with 20,480 surface elements, resulting in a grid with 3.9M elements. One advantage of this mesh is that it does not produce any polar singularity, contrary to most spherical structured meshes. For more details about the mesh design and its impact on the numerical solution, see \cite{Brchnelova2022a}.

\subsection{Discussion about input radial magnetic field}
\label{subsec:input_br}

Before analyzing the comparison with the observations, we want first to discuss the pre-processing of the synoptic maps, as it will impact the simulation results.

There are two main categories of pre-processing applied to synoptic maps for coronal simulations. PFSS-based models tend to use a Gaussian filtering, in combination with a flux-conserved remapping of the map in order to better approximate the poles \citep{Pomoell2018}. This pre-processing is important for this kind of method, since the PFSS and the subsequent WSA usually applied afterwards is very sensitive to flux distribution and expansion factor. However, for an MHD simulation, we can use another pre-processing: we can do a scale filtering by doing a spherical harmonics decomposition and selecting a maximum cut-off frequency $\ell_{max}$. This is closer to the techniques used in stellar physics, where the ZDI measurement of the magnetic field usually provides only the first 5 modes \citep{Vidotto2018}. In this study, we have chosen to apply the same pre-processing to all the maps, with an $\ell_{max}$ of 15. This is similar to a space-weather operational set-up: $\ell_{max}=15$ allows us to capture smaller structures like active regions without resolving too refined structures that would slow down the simulation. 

In all the following plots, we will divide the maps into three categories that we feel are more logical to compare. The first category are the maps in Carrington frame, which are integral maps. This category concerns WSO, HMI, GONG mrmqs and GONG mrnqs maps. All of these maps are diachronic, meaning that they are constructed by assembling observations at different times, and thus reflect only approximately the state of the solar surface at a given date.
The second category are the maps with synchronic frames with usually daily-updated data. This category concerns HMI daily, GONG mrbqs, GONG mrbqj and GONG mrzqs. These maps have a different reference frame as the 120 degrees in longitude to the left of the map are replaced with the most recently measured disk data (except for the GONG mrbqs product, which however still uses the same frame). Thus, the central meridian of the chosen date is always placed at 60 degrees from the left side of the map. Finally, we set apart the GONG-ADAPT maps, as they are 12 different variations on the same original GONG data, with just differences in parameters for the applied modeling. The selected GONG-ADAPT maps for this study are also in Carrington frame, but they are set apart because they are synchronic maps, contrary to the others which are diachronic.

\begin{figure}
    \centering
    \gridline{
        \fig{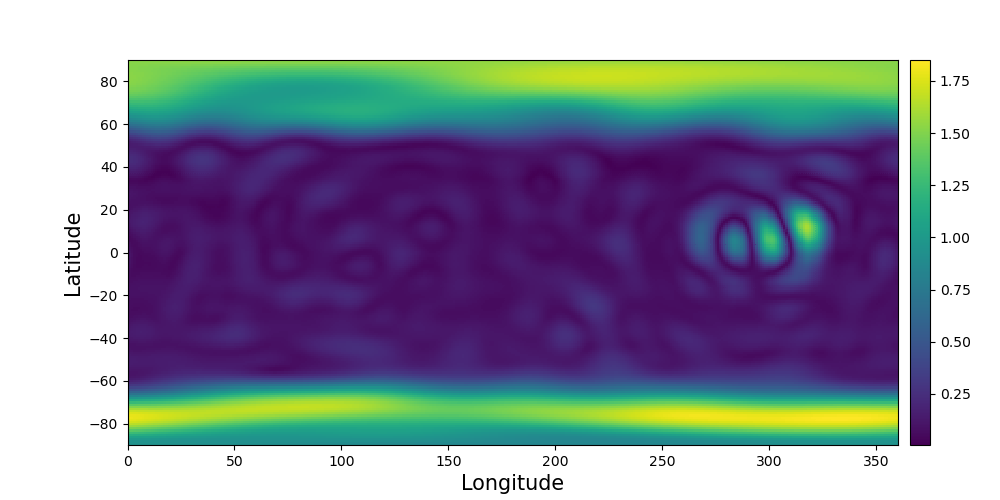}{0.49\textwidth}{(a) Carrington frame diachronic maps.}
        \fig{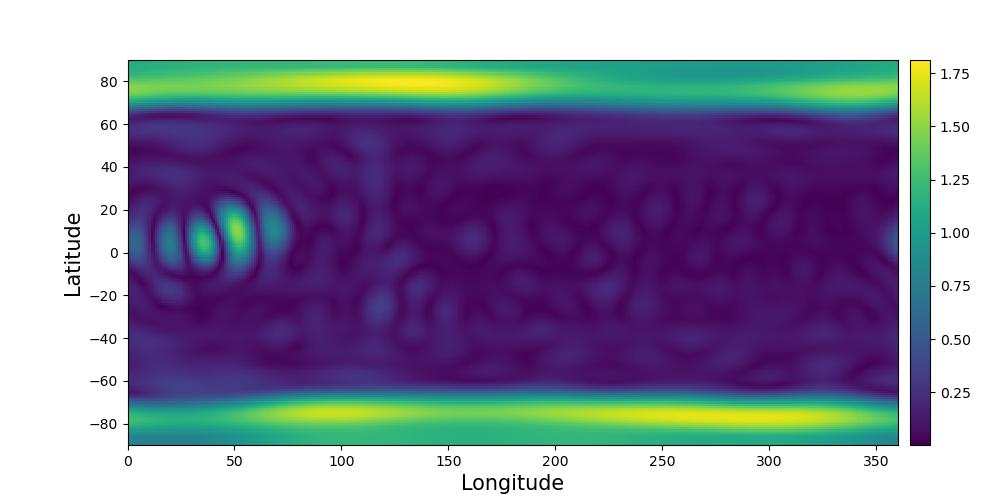}{0.49\textwidth}{(b) Synchronic frame maps.}
    }
    \gridline{\fig{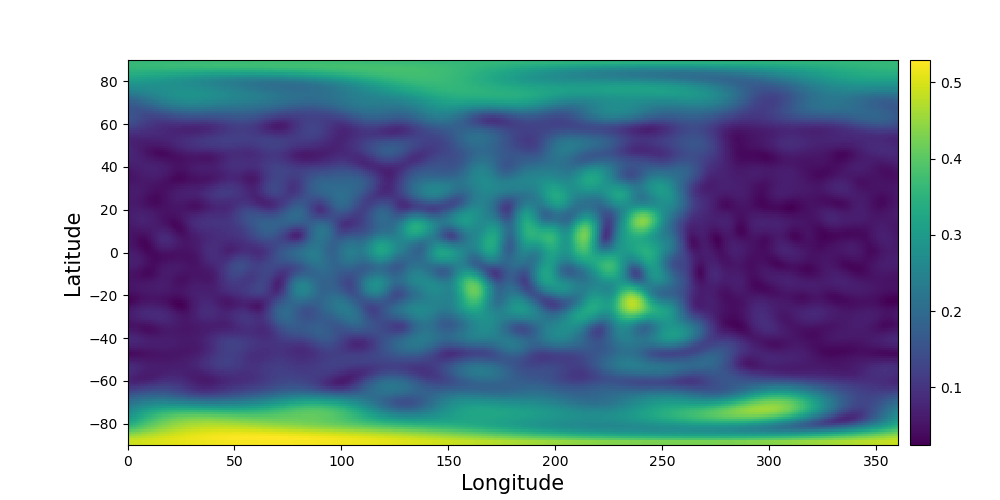}{0.49\textwidth}{(c) GONG-ADAPT realizations.}}
    \caption{Standard deviation for each pixel between input radial magnetic fields which have been derived from magnetic maps. The fields have been interpolated to the medium resolution 360x180 for comparison. The first panel shows the standard deviation from Carrington frame diachronic maps, the second one from synchronic frame maps, and the last one from all 12 GONG-ADAPT realizations for the same map. The corresponding input magnetic fields are shown in figure \ref{fig:bc_file_all}.}
    \label{fig:std_bcfile}
\end{figure}

All radial magnetic fields which have been used as boundary conditions can be found in the appendix in figure \ref{fig:bc_file_all}. The pre-processing smoothens the maps and reduces the differences due to resolution. At minimum of activity, the maps are dominated by the dipolar configuration with a positive polarity at the northern pole going down to 50 degrees in latitude, and a symmetric negative polarity at the southern pole that goes up to - 50 degrees. Despite the low activity, an active region is visible, interestingly exactly at the Carrington longitude of the date of interest (around 319 degrees). In order to show a more quantitative comparison between the boundary conditions, we display in figure \ref{fig:std_bcfile} the standard deviation as computed for all 3 above-mentioned categories for each pixel, after the input magnetic fields have been interpolated to the medium resolution of 360x180. We have chosen this resolution as it offers a good compromise between the lowest for WSO (73x30) and the highest for HMI data (3600x1440), and also because it is the most common with the chosen maps (GONG and GONG-ADAPT maps already have this resolution). The input field will anyway be interpolated to the unstructured boundary mesh, which is a bit more resolved, at the beginning of the simulation. This shows that at minimum of activity, the most significant differences between the input $B_r$ maps are located at the poles, and this for all 3 categories: above 60 degrees and below -60 degrees in latitude, Carrington frame diachronic maps have a standard deviation between 1.0 and 1.6, synchronic frame maps between 0.9 and 1.7 and GONG-ADAPT maps between 0.4 and 0.55. We can also note some other sources of differences. For Carrington frame diachronic maps (panel (a)), there is a very good agreement for the edges of the magnetic structures, but a rise of deviation at the center of the active region. This is probably due to the difference in saturation and resolution of the various maps which lead to different amplitudes of the magnetic field in the active region. The synchronic frame maps (panel (b)) also show some stronger deviation in the active region, although it is not where the maximum deviation is reached. The GONG-ADAPT maps (panel (c)) have the lowest standard deviation between the 3 categories, but they exhibit some mild deviation also at the center of the map, which is probably a result of the granulation model which is used and its various parameters that have been tested. The filling of the poles is thus going to be the main factor for explaining the differences which have been observed in the simulations.

\section{Comparing synoptic maps for the minimum of activity of July 2nd 2019}
\label{sec:comp_min}

We have selected the date of $2^{nd}$ of July 2019 because it was the most recent quiet minimum of activity date where we could combine three interesting observations in order to quantify the results of our simulations: a total solar eclipse, visible in South America at this date, provided precise white-light images of the corona, the space observatory SDO took pictures in EUV with its instrument AIA to provide maps of the coronal hole locations and the space observatory SoHO took white-light picture with its instrument LASCO to provide an estimate of the streamer belt location. Although the PSP satellite was launched by this date, it was not close to the Sun at this precise date, making it difficult to provide in-situ data in the solar corona (its closest perihelia were on $4^{th}$ April and $1^{st}$ September 2019). In this study, we will thus concentrate on remote-sensing comparisons in order to quantify the impact of the choice of the input synoptic map.

\subsection{Comparison with white-light eclipse images for streamer edges}
\label{subsec:min_wl}

\begin{figure}
    \centering
    \gridline{
        \fig{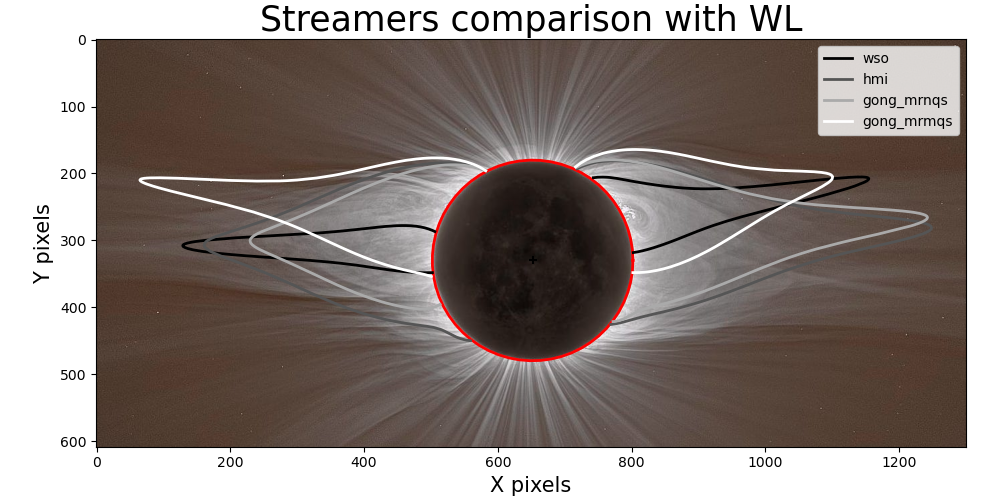}{0.49\textwidth}{(a) Carrington frame diachronic maps.}
        \fig{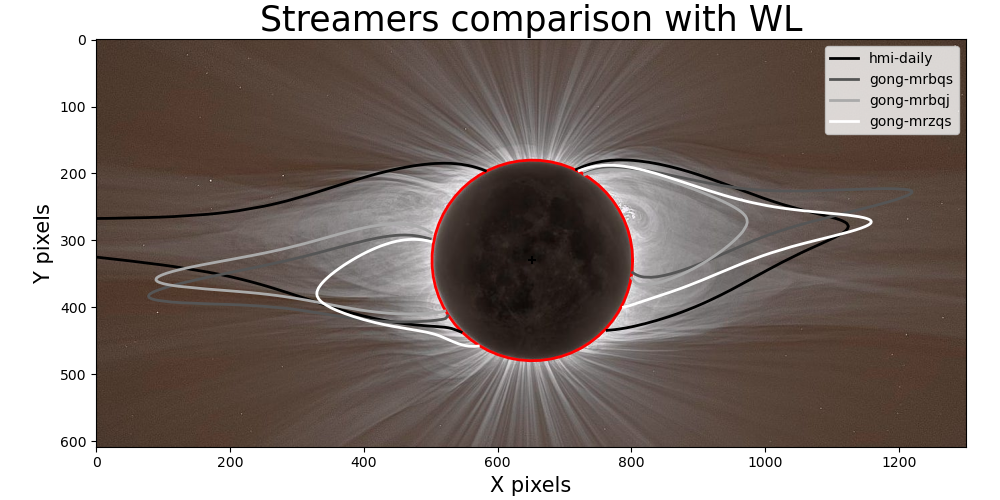}{0.49\textwidth}{(b) Synchronic frame maps.}
    }
    \gridline{\fig{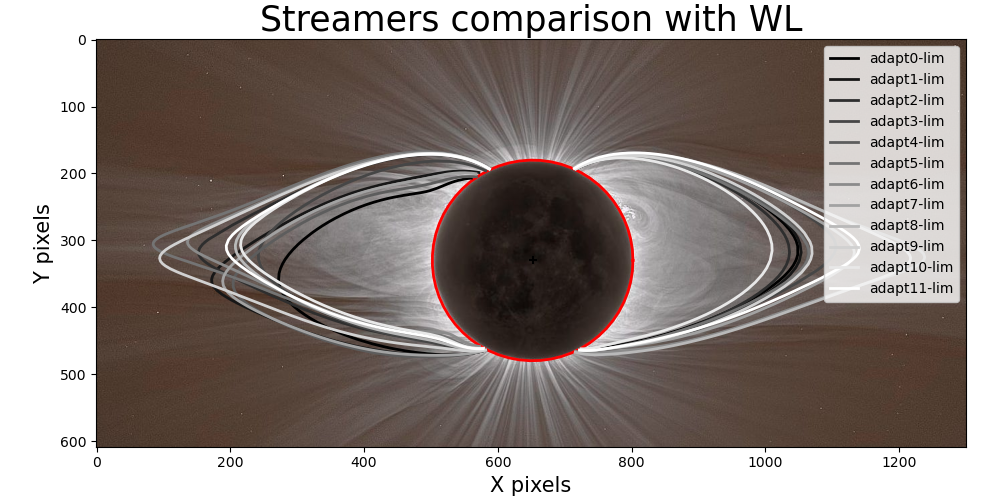}{0.49\textwidth}{(c) GONG-ADAPT realisations.}}
    \caption{Comparison of the shape of the meridional streamers with the white-light (WL) eclipse image from $2^{nd}$ of July 2019. The first panel compares the streamers from Carrington frame diachronic maps, the second one from synchronic frame maps, and the last one from all 12 GONG-ADAPT realizations for the same map. The solar disk is highlighted as a red circle as reference. Streamers contours are shown as shades of grays. All streamers have been remapped to the same size ratio using this reference and its conversion to the picture pixels, shown as axis. Credits for the WL eclipse picture: Peter Aniol, Miloslav Druckmüller.}
    \label{fig:wl_eclipse_streamers}
\end{figure}

The first comparison we show is the comparison between streamer edges and white-light eclipse images. White-light images are usually records of polarization brightness (pB) formed by Thomson scattering of photospheric light by coronal free electrons in the K corona \citep{Aschwanden2004}. Outside of solar eclipses, white-light images are generated using a coronograph from a spacecraft (e.g.\ SOHO/LASCO) or from ground-based observatories (e.g.\  COSMO/K-COR). The problem with these techniques is that the coronagraph extends above 1 solar radius, thus dimming some structures. It is actually during the solar eclipses on Earth that the solar disk is perfectly covered by the Moon, and that we can see the most precisely the shape of the streamers. For this reason, white-light pictures of eclipses have been traditionally used to constrain coronal models \citep{Mikic1999}. They are extremely useful to determine the shape of the streamers in the corona, as they reveal the underlying magnetic field structure. The white-light image we selected for $2^{nd}$ July 2019 is a composite image (128 pictures) from an open database \footnote{\url{http://www.zam.fme.vutbr.cz/~druck/Eclipse/index.htm}} maintained by Miloslav Druckmüller, that has already been used for other studies \citep{boe2020}. 

Some procedures have thus been developed to compare directly the magnetic field lines obtained from simulations with white-light pictures \cite{Wagner2022}. This is however limited to the fact that white-light images are 2D projections of the 3D configuration, which makes automatic comparisons challenging. A more quantitative approach relies upon developing a pipeline to produce artificial white-light images from simulations \citep{Mikic2018}. But this approach actually shifts the problem to the modeling of the white-light emission and the filters which are applied as post-processing for selecting the right features. In this study, we suggest another approach that tries to be both robust, so that it can be automatized, and simple enough to be implemented for all MHD models. What we do is that we compute the magnetic field lines in our simulations based on $40\times 40$ seeds which are located on a sphere at $1.01\;R_\odot$. This resolution was chosen as a good compromise between accuracy and speed. Then we select the seeds and corresponding field lines that are in the plane perpendicular to the observer line of sight at the date of the event. From these we can extract the largest closed magnetic field line, which corresponds then to the edge of the streamers as seen from the Earth. We can finally superpose these edges on the white-light images, by projecting the field lines in the 2D plane and adjusting them to the size of the picture (the reference is the radius of the solar disk where we find the conversion between physical and pixel size). The entire procedure is completely automatic and operated by Python scripts.

The results are shown in figure \ref{fig:wl_eclipse_streamers}. As stated before, we divide the simulations into 3 categories based on the frame of the maps (Carrington frame diachronic, synchronic frame and GONG-ADAPT realizations). For each subgroup, we show the white-light image in gray scale in the background to enhance the features. On top of it, we show the solar disk edge as a red circle. This feature is important because it is actually detected automatically using hysteresis thresholding, and used to adjust the size of the streamers from the simulation to the eclipse picture. Finally, we plot the streamer edges extracted from each simulation in shades of gray. We note that for this date, the streamers are remarkably large, as shown by the white-light image. We can distinguish by eye one streamer on the left, and two streamers on the right that overlap, which probably means that they are not located at the same longitude. At the poles, we can clearly see open magnetic field lines that are almost vertical. This is typical of a minimum of activity configuration. The size of the streamers and the complexity of the structures visible between 1 and 1.5 solar radii indicate that they may be overarching pseudo-streamers rather than helmet streamers \citep{Wang2007}. These structures are still the most relevant as they indicate the limit between closed and open magnetic field lines. Inside each subgroup, we can already see a wide variety of results. For the Carrington frame diachronic maps, the HMI and GONG mrnqs runs yield very good results, but the two other simulations are completely off. The WSO streamers are way too thin, while the GONG mrmqs streamers are shifted upwards to a position that does not match anymore the white-light image. This is not surprising, because the GONG mrnqs is supposed to be more accurate than the GONG mrmqs thanks to its zero-point correction. For the synchronic frame maps, the best result is given by the HMI run, although the left streamer is too big (5 solar radii instead of 3.5). Between the GONG cases, the best result is given by GONG mrzqs, although the left streamer is too small and shifted too downwards. The difference between GONG mrbqs and GONG mrbqj is minimal, with just the right streamers having a better size with GONG mrbqs. This is what we would have expected, since the GONG mrzqs is the most accurate and physical map. It is however surprising that our model performs less efficiently with the synchronic frame maps than with the Carrington frame diachronic maps which have a bigger asynchronicity between the data. For the GONG-ADAPT runs, there is a bigger diversity in the results that what could have been expected based on the standard deviation study, with the left streamer edge ranging from 2.5 to 3.5 solar radii, and the right streamer from 2 to 4 solar radii. The overall agreement is still very good, although it is clearly visible that some realizations yield better simulations than others. All results are summed up in a more quantitative way in table \ref{tab:metrics} (see section \ref{subsec:map_scores} for the corresponding discussion).

\subsection{Comparison with EUV images for coronal hole boundaries}
\label{subsec:min_ch}

\begin{figure}
    \centering
    \gridline{
        \fig{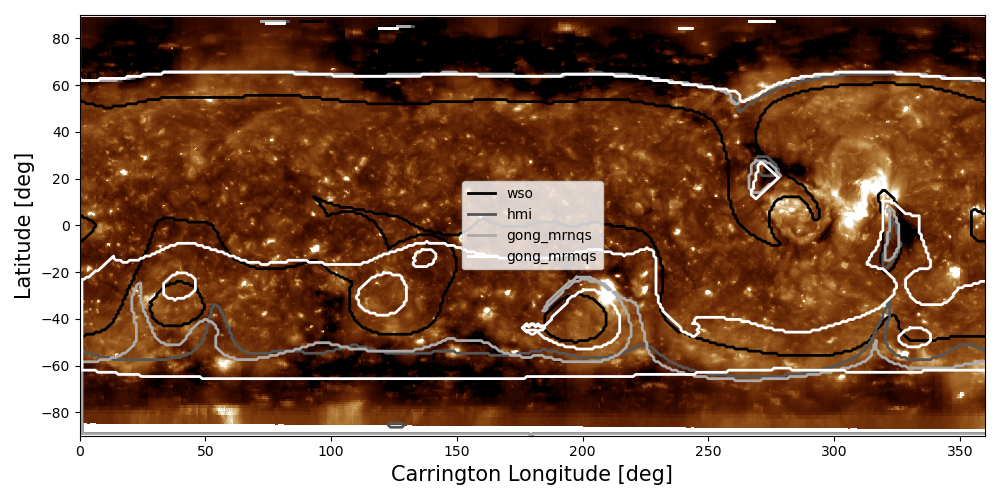}{0.49\textwidth}{(a) Carrington frame diachronic maps.}
        \fig{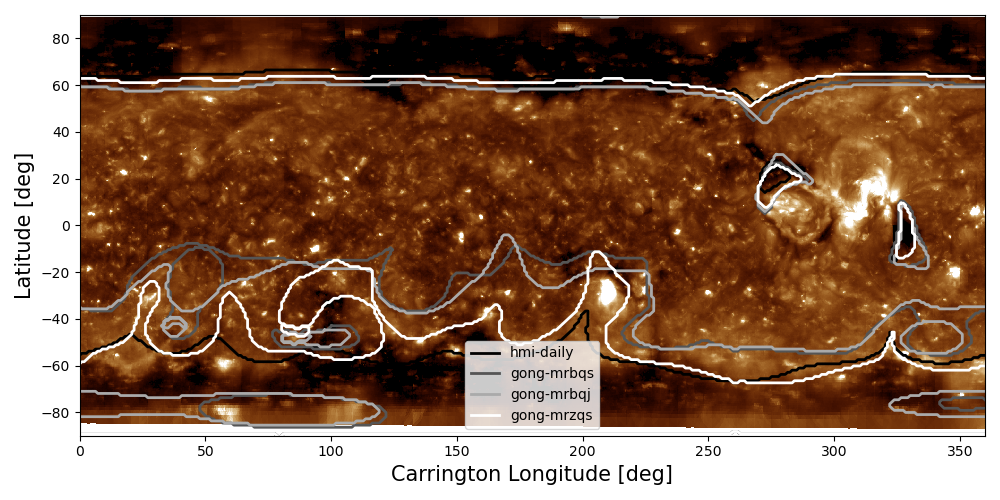}{0.49\textwidth}{(b) Synchronic frame maps.}
    }
    \gridline{\fig{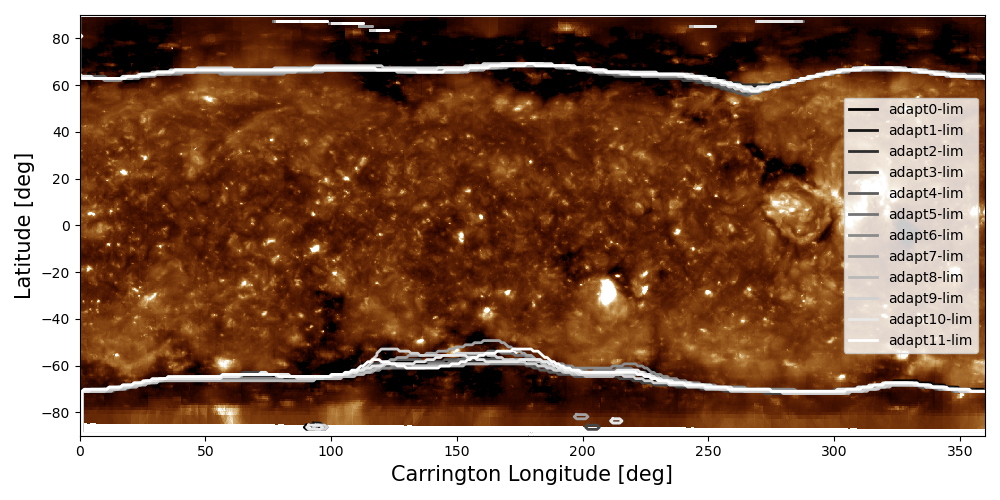}{0.49\textwidth}{(c) GONG-ADAPT realisations.}}
    \caption{Comparison of the contours of the coronal holes (CH) with the EUV synoptic map from Carrington rotation 2219 from SDO/AIA (channel 195). The first panel compares the streamers from Carrington frame diachronic maps, the second one from synchronic frame maps and the last one from all 12 GONG-ADAPT realizations for the same map. Coronal hole contours from simulations are shown as shades of grays.}
    \label{fig:euv_ch}
\end{figure}

The second physical quantity we use for comparison is the EUV emission at 195 \AA, which is the wavelength recommended to automatically extract coronal hole boundaries \citep{Wagner2022, Badman2022}. Coronal holes are dimmings in the EUV emission, which corresponds to regions of open magnetic field lines associated with cooler plasma \citep{Cranmer2009}. The synoptic map we use is from the official SDO/AIA website and consists of a reconstruction of the full solar disk based on daily data, following the same principle as the HMI magnetic maps. It has also been remapped to latitude coordinates, which can create some artifacts at the poles due to the line of sight constraints. Again, artificial EUV emissions can be generated from simulations to provide an accurate comparison \citep{Lionello2009, Parenti2022}. The polytropic approximation we use for the coronal heating does not allow us to use such techniques, but we have access to the information about the open magnetic field lines in the simulation. We then proceed to find the boundaries between closed and open field lines at the surface of the star, using a sphere of 400x200 seed points at $1.01\;R_\odot$. We follow the field lines to see if they reach the end of the computational domain at $20\;R_\odot$: if they do, they are open field lines, if not, they are closed field lines. This allows us to retrieve contours of the open field line regions at the surface of the star, that we can directly compare with the coronal hole synoptic map. This is not completely a direct comparison, as the EUV emission corresponds to the photosphere, while the wind simulations start at the lower corona above the transition region, but we do not have measurements at this height, and we assume that the change of structure in the coronal hole is minimal over this interval. Similar comparisons have been performed in previous studies with positive results \citep{Badman2022}.

We plot the results in figure \ref{fig:euv_ch}. For each subgroup of map, we over-plot the contours obtained from the various simulations on the synoptic EUV map. At the chosen date, there are mostly polar coronal holes in dark, and also several dimmer equatorial coronal holes at 220, 270 and 330 degrees in longitude. The contours from the simulations have to match as closely as possible the contours of these dark regions. For the Carrington frame diachronic maps, we can see that most of the simulations cover reasonably the northern coronal hole, except for the WSO map which has an incursion towards the equator at 270 degrees in longitude, which is not visible in the EUV data. The HMI and GONG mrnqs simulations capture well the southern and equatorial coronal holes, but the WSO and GONG mrmqs both overestimate them. Once again, this is not surprising due to the fact that GONG mrnqs is the corrected version of the GONG mrmqs map. For the synchronic frame maps, we observe that the northern and equatorial coronal holes are well captured. The best results for the southern coronal hole are given by the HMI simulation, while all the GONG simulations tend to overestimate it. We can see however the effect of the correction in the GONG mrzqs map, since it is the only GONG map not to exhibit closed field lines at the southern pole. For the GONG-ADAPT simulations, there is little to no disagreement between the different realizations, although the southern coronal hole is still the one with the most differences. The agreement is very good for both polar coronal holes, but all realizations completely miss the equatorial coronal holes, which is surprising given the accuracy of the models used and the fact that other maps capture them with the same pre-processing. All results are summed up in a more quantitative way in table \ref{tab:metrics} (see section \ref{subsec:map_scores} for the corresponding discussion).

\subsection{Comparison with white-light coronagraphs for streamer belt}
\label{subsec:min_hcs}

\begin{figure}
    \centering
    \gridline{
        \fig{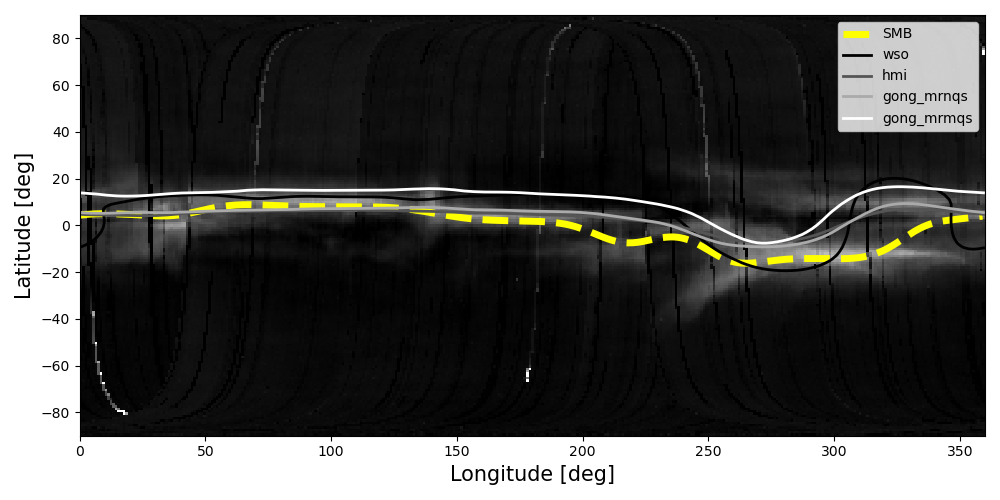}{0.49\textwidth}{(a) Carrington frame diachronic maps.}
        \fig{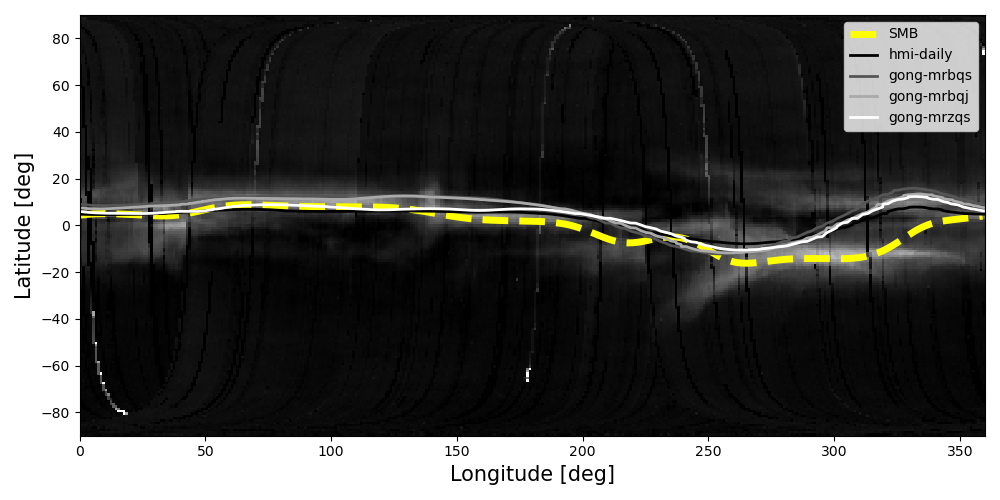}{0.49\textwidth}{(b) Synchronic frame maps.}
    }
    \gridline{\fig{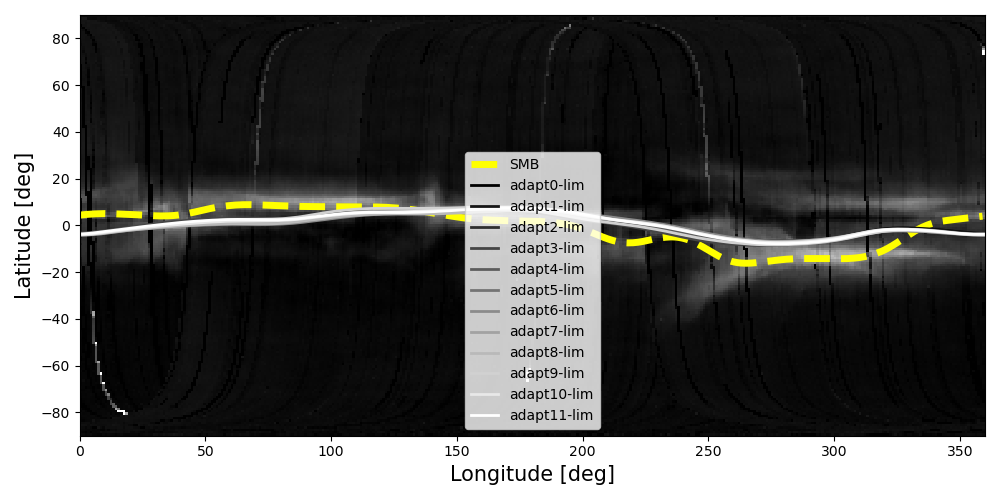}{0.49\textwidth}{(c) GONG-ADAPT realisations.}}
    \caption{Comparison of the shape of the streamer belt (SB) with the white-light synoptic maps from 2nd of July 2019 from SoHO/LASCO/C2. The first panel compares the streamers from Carrington frame diachronic maps, the second one from synchronic frame maps and the last one from all 12 GONG-ADAPT realizations for the same map. The SMB line inferred from observations is in yellow and dashed line, while the current sheet inferred from simulations is in shades of grays. Credits for the SMB maps: Nicolas Poirier (IRAP).}
    \label{fig:hcs_wl_5rs}
\end{figure}

The last comparison with observational data we want to make is the comparison between the white-light streamer belt and the heliospheric current sheet. The coronagraph LASCO C2 aboard SoHO capture white-light images between 1.5 and 6 solar radii. This data can then be assembled as a synoptic map over a Carrington rotation to give an estimate of the streamer belt (SB), which can be assumed to host the heliospheric current sheet (HCS) and act then as a proxy for it at around $5\;R_\odot$ \citep{Poirier2021}. From the simulations, it is easy to directly extract the HCS, as it is the separation between the positive and negative polarity of the radial magnetic field in the computational domain. Once again, this method has already been used in previous studies with positive results \citep{Badman2022}.

We plot the results in figure \ref{fig:hcs_wl_5rs}. The background shows the white-light synoptic maps in gray scale, with the SB highlighted with a yellow dashed line. Because we are looking at a minimum of activity, the HCS is very flat as the current sheet is almost horizontal, with a slight deviation between 250 and 330 degrees in longitude that is due to the active region discussed before. The HCS extracted from simulations is plotted as a line in gray scale. For the Carrington frame diachronic runs, we see once again that the HMI and GONG mrnqs simulations yield the best result, although the gap between 250 and 330 degrees in longitude seems more difficult to capture, most probably because of the active region located at this exact spot. The WSO and GONG mrmqs simulations show a shift upwards compared to the actual SB, and the WSO simulation shows the biggest deviation between 300 and 360 degrees in longitude. For the synchronic frame maps, most of the simulations agree very well, with just a slight overestimation of the SB by the GONG mrbqj simulation. For the GONG-ADAPT realizations, there is also little to no variation between all the various simulations, that capture the SB quite well. The better agreement between simulations can be explained by the fact that this quantity is observed at $5\;R_\odot$, a distance at which the magnetic field is more uniform. All results are summed up in a more quantitative way in table \ref{tab:metrics} (see section \ref{subsec:map_scores} for the corresponding discussion).

\section{Discussion for space-weather applications}
\label{sec:discussion}

\subsection{Assessing the impact for space weather forecasting} \label{subsec:min_space_weather_forecast}

\begin{figure}
    \centering
    \gridline{
        \fig{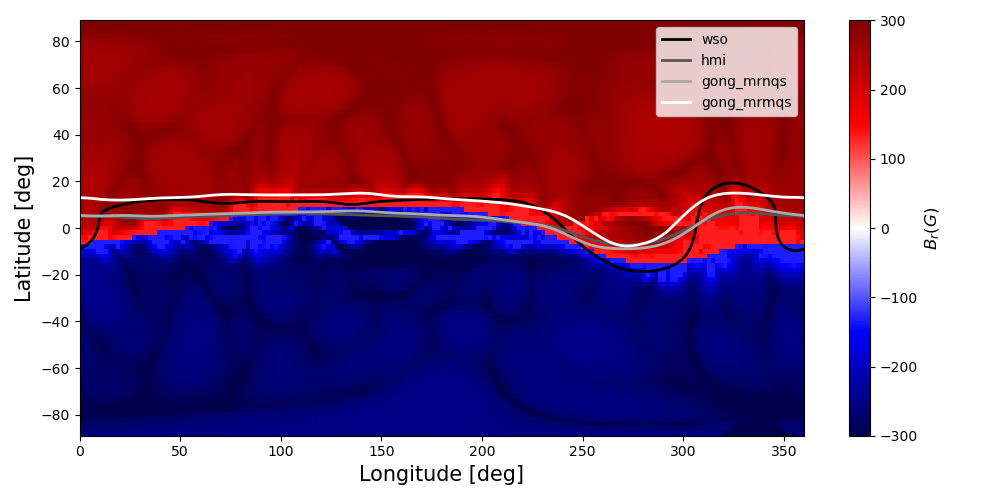}{0.49\textwidth}{(a) Carrington frame diachronic maps.}
        \fig{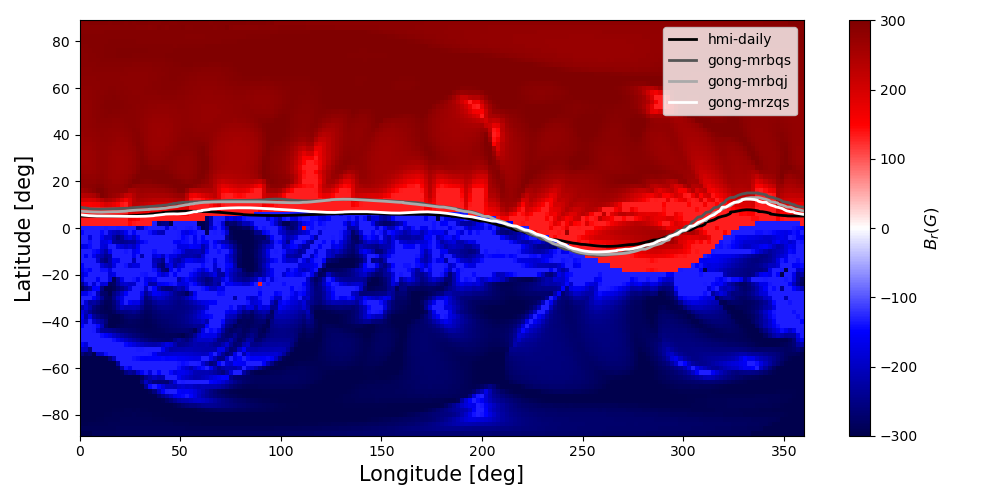}{0.49\textwidth}{(b) Synchronic frame maps.}
    }
    \gridline{\fig{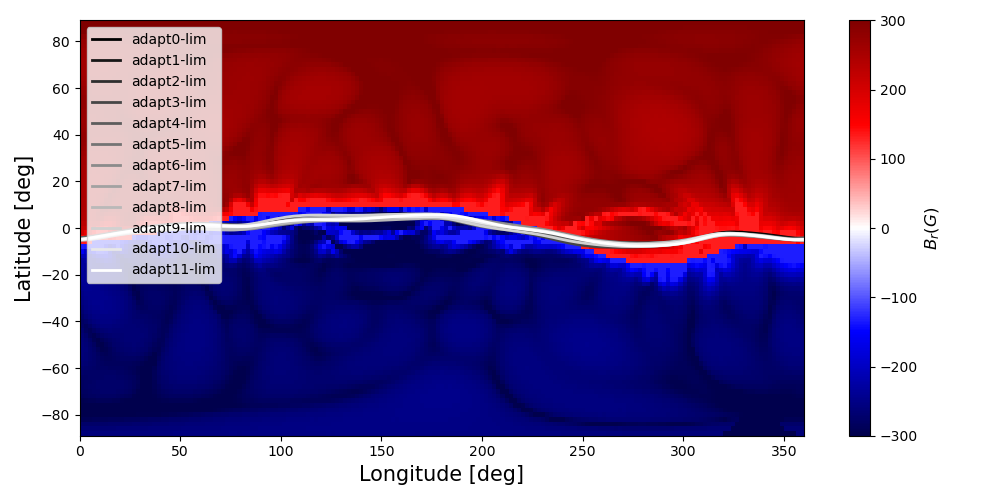}{0.49\textwidth}{(c) GONG-ADAPT realisations.}}
    \caption{Comparison between the typical HCS extrapolated by a PFSS+SCS method at 0.1 AU and the ones extracted from our MHD simulations. The first panel compares the HCS from Carrington frame diachronic maps, the second one from synchronic frame maps and the last one from all 12 GONG-ADAPT realizations for the same map. The background color shows the radial magnetic field $B_r$ polarity for the extrapolation (red for positive, blue for negative). For panels (a) and (c), the PFSS is based on a GONG-ADAPT map to provide a Carrington-rotation frame. For panel (b), it is based on a GONG mrbqs map to provide a synchronic frame.}
    \label{fig:hcs_wsa_01au}
\end{figure}

In an operational set-up for space weather forecast, the coronal part of the model chain is useful for providing the physical quantities at around $20\;R_\odot$ to heliospheric propagators that can compute them all the way to Earth. Currently, in operational environments the coronal part is handled through semi-empirical extrapolations, such as the WSA model combined with PFSS and SCS for the magnetic field part \citep{Pomoell2018}. This is due to the fact that current MHD models are too slow to be used in an operational context, although it has been demonstrated on numerous occasions that they are more accurate \citep{Samara2021}. This is a limitation that our code does not have, thanks to its implicit solving method \citep{Perri2022}. It is then interesting to wonder what are the differences we would observe if we were to couple our MHD model to EUHFORIA for example, and see the modifications at this interface. As we use a polytropic version of the code for now, thus it is not interesting to do the coupling all the way to Earth, because we already know it will not compare well with in-situ measurements at L1. However, we can already compare to typical forecasts. Our velocity, density and temperature are also going to be limited by the polytropic assumption, so for the moment the best quantity to compare is the radial magnetic field $B_r$.

We plot the results in figure \ref{fig:hcs_wsa_01au}. The background color shows the radial magnetic field $B_r$ extrapolated at 0.1 AU by PFSS+SCS. The positive polarity is shown in red, negative polarity in blue, and the HCS is located at the border between the two. For panels (a) and (c), the PFSS extrapolation is based on the realization 1 from GONG-ADAPT to provide the right frame. At the moment, the prevision models do not offer other maps to work with. For panel (b), it is based on a GONG mrbqs map to have the synchronic frame. We over-plot the HCS extracted from our MHD simulations around 0.1 AU for comparison. We can see that compared to the HCS at $5\;R_\odot$, the HCS at 0.1 AU is not very different, as the global geometry of the magnetic field is already fixed at this distance. We can see however a significant deviation from the HCS from the WSA model. This is surprising for synchronic frame maps, since they are based on exactly the same map for GONG mrbqs, only the model changes. From empirical to MHD, we can see that the gap around the active region is accentuated for the PFSS extrapolation. For the GONG-ADAPT realizations, the MHD model also tends to reduce the north-south variations and flatten the HCS. This is important for space weather forecasts, as a difference of several tens of degrees at 0.1 AU will increase even further and become even more significant at 1 AU. It is well known that a southwards inclined IMF $B_z$ for CMEs leads to more geo-effective intense magnetic storms, which means that these difference will have a significant impact on forecasts at Earth \citep{Balan2014cme}. 

\subsection{Which map should we choose?}
\label{subsec:map_scores}

\begin{table}[]
    \centering
    \begin{tabular}{|c||c|c|c|c|}
        \hline
        Map & Streamers ratio & Polar CH ratio & Eq. CH ratio & SB deviation \\ \hline \hline
        WSO & \colorbox{red}{left: 28.0\%}, \colorbox{red}{right: 24.0\%} & \colorbox{red}{North: 72.8\%}, \colorbox{yellow}{South: 33.7\%} & \colorbox{orange}{10.7\%} & \colorbox{red}{$\delta_{max}=30.8\degree$}, \colorbox{orange}{$\delta_{mean}=9.22\degree$} \\ \hline
        HMI & \colorbox{lime}{left: 84.2\%}, right: 74.7\% & North: 86.1\%, South: 40.6\% & \colorbox{green}{37.4\%} & $\delta_{max}=17.5\degree$, $\delta_{mean}=4.88\degree$ \\ \hline
        GONG (mrmqs) & left: 54.4\%, \colorbox{yellow}{right: 37.7\%} & North: 87.1\%, \colorbox{red}{South: 23.9\%} & 8.8\% & \colorbox{orange}{$\delta_{max}=27.9\degree$}, \colorbox{red}{$\delta_{mean}=11.9\degree$} \\ \hline
        GONG (mrnqs) & left: 74.9\%, right: 65.6\% & North: 86.2\%, South: 42.0\% & \colorbox{teal}{26.2\%} & $\delta_{max}=19.1\degree$, $\delta_{mean}=4.98\degree$ \\ \hline
        HMI (sync.) & left: 66.2\%, right: 70.1\% & North: 86.3\%, South: 40.1\% & \colorbox{lime}{65.5}\% & $\delta_{max}=16.1\degree$, \colorbox{lime}{$\delta_{mean}=4.30\degree$} \\ \hline
        GONG (mrbqs) & \colorbox{yellow}{left: 39.1\%}, right: 41.6\% & \colorbox{yellow}{North: 80.3\%}, South: 33.9\% & 11.6\% & \colorbox{yellow}{$\delta_{max}=23.9\degree$}, \colorbox{yellow}{$\delta_{mean}=7.35\degree$} \\ \hline
        GONG (mrbqj) & left: 47.3\%, \colorbox{orange}{right: 32.8\%} & \colorbox{orange}{North: 79.2\%}, \colorbox{orange}{South: 32.5\%} & \colorbox{yellow}{11.4\%} & $\delta_{max}=20.5\degree$, $\delta_{mean}=6.43\degree$ \\ \hline
        GONG (mrzqs) & \colorbox{orange}{left: 29.1\%,} right: 53.6\% & North: 85.2\%, South: 39.2\% & 20.4\% & $\delta_{max}=19.7\degree$, \colorbox{green}{$\delta_{mean}=4.66\degree$} \\ \hline
        ADAPT (1) & left: 64.3\%, right: 77.8\% & \colorbox{green}{North: 88.1\%}, South: 44.4\% & \colorbox{red}{0.0\%} & $\delta_{max}=10.5\degree$, $\delta_{mean}=5.36\degree$ \\ \hline
        ADAPT (2) & left: 61.7\%, right: 77.1\% & North: 87.9\%, South: 44.1\% & \colorbox{red}{0.0\%} & $\delta_{max}=9.99\degree$, $\delta_{mean}=5.60\degree$  \\ \hline
        ADAPT (3) & left: 69.4\%, right: 72.4\% & \colorbox{lime}{North: 88.3\%}, South: 44.0\% & \colorbox{red}{0.0\%} & $\delta_{max}=10.5\degree$, $\delta_{mean}=5.57\degree$ \\ \hline
        ADAPT (4) & \colorbox{teal}{left: 77.0\%}, \colorbox{teal}{right: 85.5\%} & North: 87.9\%, South: 43.9\% & \colorbox{red}{0.0\%} & \colorbox{teal}{$\delta_{max}=9.69\degree$}, \colorbox{teal}{$\delta_{mean}=4.76\degree$} \\ \hline
        ADAPT (5) & left: 61.4\%, right: 79.5\% & North: 87.8\%, South: 44.5\% & \colorbox{red}{0.0\%} & $\delta_{max}=9.84\degree$, $\delta_{mean}=5.09\degree$ \\ \hline
        ADAPT (6) & left: 66.3\%, right: 78.1\% & North: 87.5\%, South: 44.1\% & \colorbox{red}{0.0\%} & $\delta_{max}=10.0\degree$, $\delta_{mean}=5.84\degree$ \\ \hline
        ADAPT (7) & left: 72.1\%, right: 78.5\% & North: 87.2\%, South: 43.6\% & \colorbox{red}{0.0\%} & $\delta_{max}=10.4\degree$, $\delta_{mean}=6.20\degree$ \\ \hline
        ADAPT (8) & left: 61.9\%, \colorbox{lime}{right: 87.9\%} & North: 87.4\%, \colorbox{lime}{South: 45.3\%} & \colorbox{red}{0.0\%} & \colorbox{green}{$\delta_{max}=9.63\degree$}, $\delta_{mean}=5.75\degree$ \\ \hline
        ADAPT (9) & left: 75.4\%, right: 77.6\% & North: 87.7\%, South: 43.4\% & \colorbox{red}{0.0\%} & $\delta_{max}=10.3\degree$, $\delta_{mean}=5.91\degree$ \\ \hline
        ADAPT (10) & left: 61.3\%, right: 80.5\% & \colorbox{teal}{North: 88.0\%}, \colorbox{green}{South: 44.9\%} & \colorbox{red}{0.0\%} & \colorbox{lime}{$\delta_{max}=9.39\degree$}, $\delta_{mean}=4.99\degree$ \\ \hline
        ADAPT (11) & \colorbox{green}{left: 80.0\%}, right: 64.1\% & \colorbox{green}{North: 88.1\%}, \colorbox{teal}{South: 44.7\%} & \colorbox{red}{0.0\%} & $\delta_{max}=10.4\degree$, $\delta_{mean}=5.73\degree$ \\ \hline
        ADAPT (12) & left: 76.1\%, \colorbox{green}{right: 85.8\%} & North: 87.9\%, South: 44.5\% & \colorbox{red}{0.0\%} & $\delta_{max}=10.0\degree$, $\delta_{mean}=5.52\degree$ \\ \hline
    \end{tabular}
    \caption{Summary of the comparison between COCONUT MHD simulations of July $2^{nd}$ 2019 based on various magnetic maps and the available observational data. We use 3 quantitative metrics to evaluate the maps: we compute the percentage of overlap between the streamers edges, the percentage of coverage between the polar and equatorial coronal holes, and the mean and maximum angle of deviation between the SB and the HCS. For more details on the metrics, see appendix \ref{app:metrics}. The best result for each metrics is highlighted in \colorbox{lime}{lime}, the second best in \colorbox{green}{green} and the third best in \colorbox{teal}{deep green}. The worst result for each metrics is highlighted in \colorbox{red}{red}, the second worse in \colorbox{orange}{orange} and the third worse in \colorbox{yellow}{yellow}. The sync. abbreviation stands for "synchronic frame".}
    \label{tab:metrics}
\end{table}

Based on the previous results, we have summarized all our results in table \ref{tab:metrics} to create a scoreboard of all the studied maps for this given date with the COCONUT model.

In order to be able to be more quantitative, we have used 3 metrics based on the comparisons described in the previous section. At first, in order to better compare the streamers' edges, we have computed the percentage of overlap between the observations and the simulations. From the white-light eclipse picture, we have extracted a visual estimation of the streamers' edges in the plane perpendicular to the observer's line of sight. 
This method is of course limited by the fact that the white-light picture without post-processing offers only a 2D projection of the 3D structure of the streamers.
We have then identified all the points that are inside the selected contour as belonging to the streamer, and have plotted the defined surface along with the streamer from the simulation. We then compute the ratio between the number of pixels that belong to the two streamers (the one from the eclipse picture and the one from the simulation) and the number of pixels within the biggest streamer of the two. This way of computing the ratio allows us to not give a perfect score to simulation streamers that are bigger than the observation streamer and would then include it. The corresponding maps for the computation of this ratio can be found in figure \ref{fig:streamers_comp_all} which provides a visual representation. Then, for the coronal holes comparison, we have used a similar technique of area ratio. We have extracted the pixels that belong to the coronal holes from the EUV synoptic map by applying the EZSEG algorithm developed by Predictive Science Inc. \citep{Caplan2016ApJ}. The software is available as part of the EUV2CHM Matlab package from the Predictive Science Inc. website \footnote{\url{https://www.predsci.com/chd/}}. We have converted the algorithm in Python to be able to use directly on our pipeline for the EUV synoptic map. This
algorithm uses an initial intensity threshold to acquire
coronal hole locations in an EUV image, and then
uses an area growing technique to define connected regions. This continues until a second intensity threshold is reached, or the condition for connectivity is not met. The dual-thresholds and connectivity conditions (essentially the number of consecutive pixels), are defined on input. We experimented with the optimal input parameters, and found that for this map the best result was obtained with a connectivity of 3 neighbors, a first threshold at 20 and a second threshold at 35. The coronal hole for the simulations were determined as said before by using seeds for the field lines and checking whether the field line would reach the outer boundary of the computational domain. We then computed the ratio of the number of pixels in both coronal hole detections to the number of pixels from the coronal holes from the simulation. That way, this percentage represents how accurate the coronal hole from the simulation is. We have separated polar and equatorial coronal holes by defining the equatorial region as being between -40 and 40 degrees in latitude. The corresponding maps for the computation of this ratio can be found in figure \ref{fig:ch_comp_all} to have a visual representation. Finally, we compute the deviation of the HCS from the SB. In order to do so, we interpolate the two lines at the same resolution, and compute for each longitude the difference in latitude in degrees. We then process the results to compute the maximum and mean deviation for each map.

Table \ref{tab:metrics} gives thus an overview of the quality of the maps for this specific date in combination with the current standard set-up of the COCONUT model (described in section \ref{sec:cf}). What we see is that the GONG-ADAPT runs yield very good results for the streamers, the polar coronal holes and the HCS, but completely fail to capture the equatorial coronal holes. This may be due to the fact that the coronal holes were quite small, but other simulations with different maps managed to capture them with good accuracy with the same pre-processing. This then means that this pre-processing does not work well for our use of the GONG-ADAPT maps with the COCONUT model, and thus should be adapted for these maps. This is important information for forecasts, as equatorial coronal holes are often the sources of high-speed streams that are going to reach Earth and can cause mild space weather events. The other category of runs that score well are the ones base on the HMI maps, both Carrington frame diachronic and synchronic frame. Contrary to the GONG-ADAPT simulations, they have a high score for the equatorial coronal holes, and they manage to score high in almost all the metrics. For the GONG runs, the results are overall pretty unsatisfactory, especially for the GONG mrbqs and mrbqj, which is surprising because these are the most used maps for forecasting. This table gives us some useful guidelines in order to use the COCONUT code for space-weather applications in the most efficient way. From the table, it seems clear that the only acceptable synchronic frame map we could use with COCONUT is the GONG mrzqs. Same thing for the Carrington frame diachronic maps, the correction for the GONG mrnqs map really improves the quality of the simulation. Finally, the WSO runs score the worst in almost all the metrics, and are thus not recommended to use as it is with our code. They can however be adjusted with a more elaborate and custom pre-processing, but it is not clear whether this is applicable to space weather forecasting \citep{Samara2021}.

To conclude, the runs that agree with most of the metrics are the ones based on the GONG-ADAPT maps, although they may require additional pre-processing in order to better treat the equatorial coronal holes. The second-best choices that score good on average are the HMI simulations, both Carrington frame diachronic and synchronic frame. This may be actually the best choice for operational previsions with COCONUT, and yet to this date few data centers have tested them with other models in operational set-ups. Instead, the second choice is usually the GONG mrbqs map. For our model, it scores relatively bad (third-worse). A better choice would be for us the GONG mrzqs for the synchronic frame maps and the GONG mrmqs for the Carrington frame diachronic maps. To this date, not all prevision centers use the zero-corrected GONG maps, which appears to be better suited since they were designed to provide better results for the solar poles. These conclusions are of course tied to the date and model that we used, and would need a more extensive statistical study to be generalized. It is however likely that for the same approximations (ideal MHD and polytropic heating) and similar boundary conditions, other models would find similar results. It would also be interesting to see if the same conclusion holds for a maximum of activity configuration, which would probably show even more disparities between the maps \citep{Yeates2018}. Finally, this is based on remote-sensing coronal validation, and should also be confronted with in-situ heliospheric metrics to have a complete view of the impact for space weather forecasting, but this requires a better description of the coronal heating that we leave for future work.

\subsection{Do solar poles matter for space weather?}
\label{subsec:poles}

\begin{figure}
    \centering
    \gridline{
        \fig{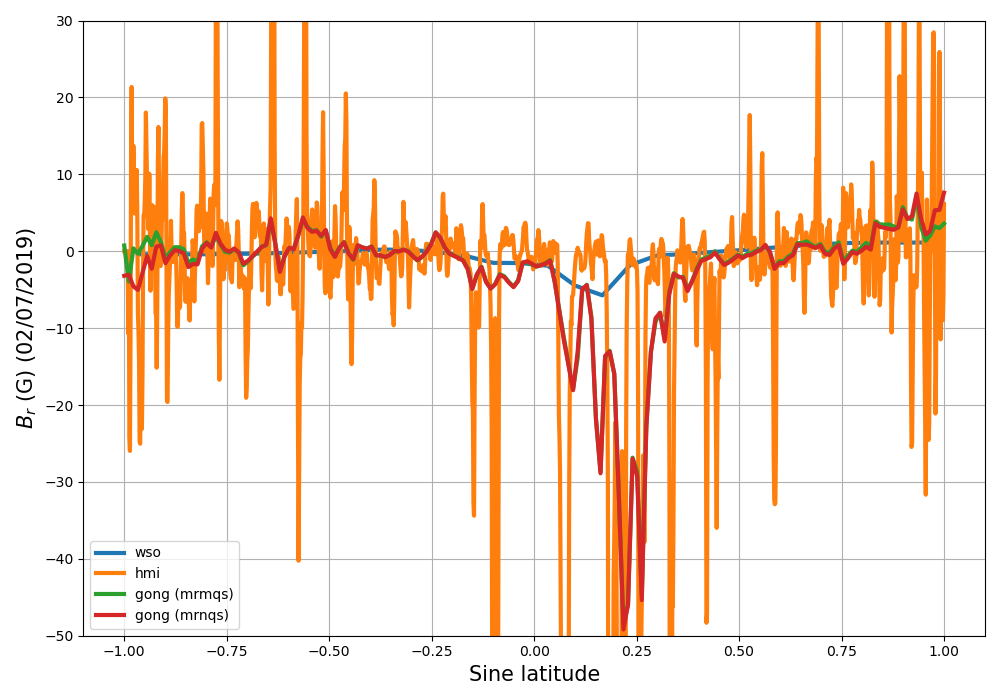}{0.49\textwidth}{(a) Carrington frame diachronic original maps.}
        \fig{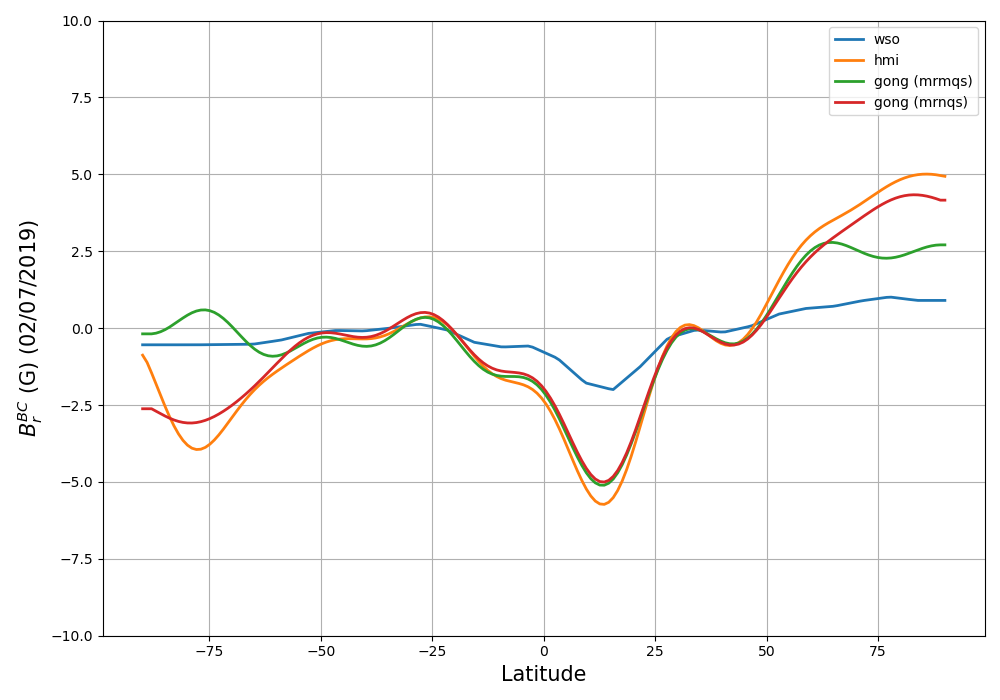}{0.49\textwidth}{(b) Carrington frame diachronic pre-processed maps.}
    }
    \gridline{
        \fig{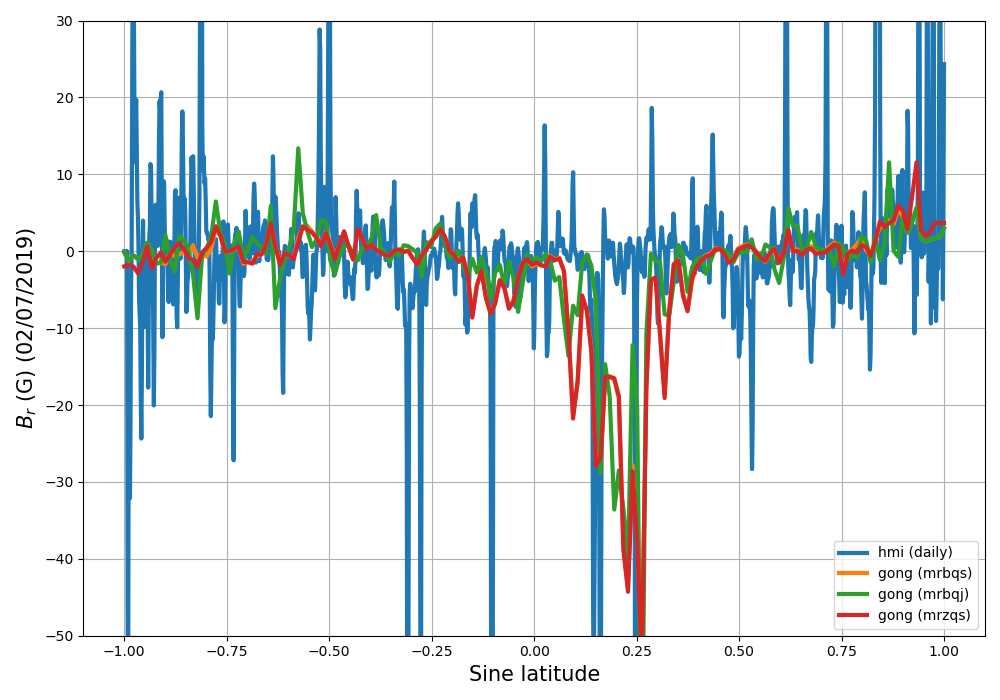}{0.49\textwidth}{(c) Synchronic frame original maps.}
        \fig{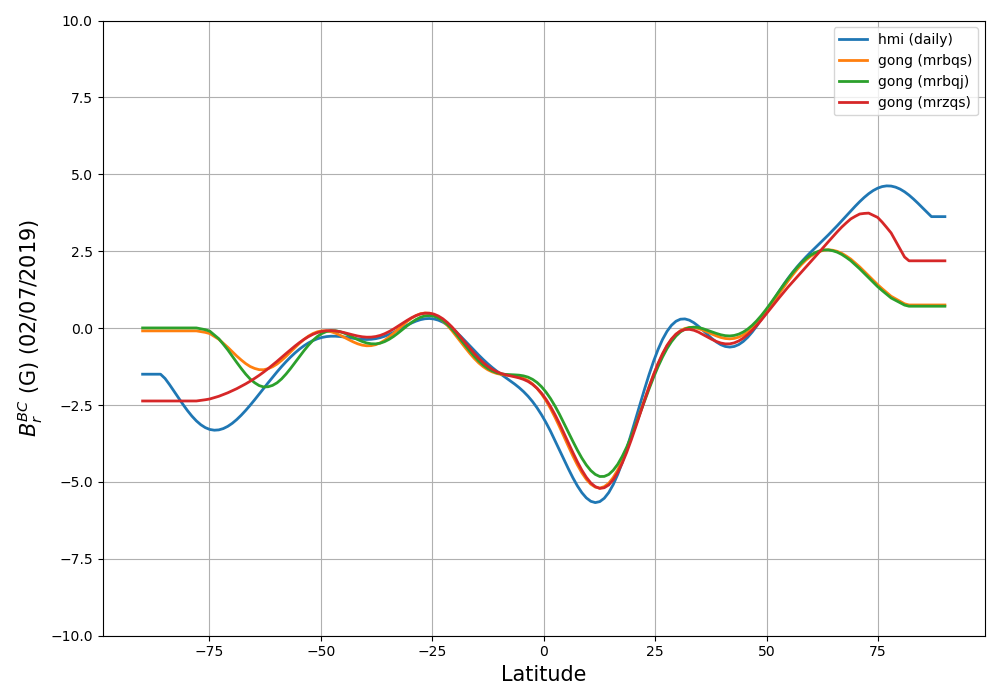}{0.49\textwidth}{(d) Synchronic frame pre-processed maps.}
    }
    \gridline{
        \fig{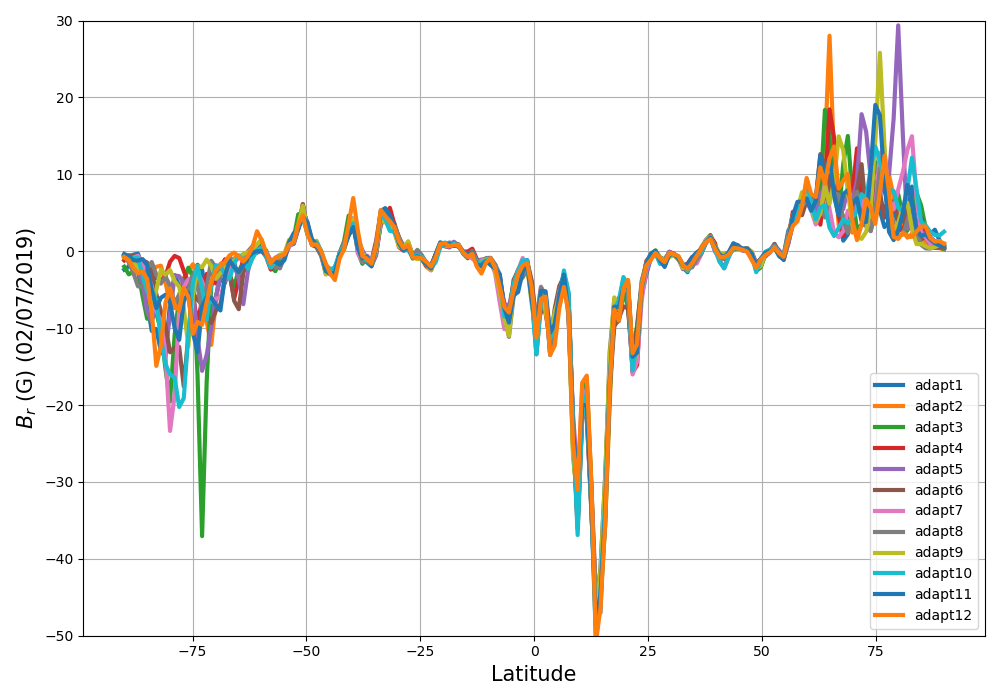}{0.49\textwidth}{(e) GONG-ADAPT original maps.}
        \fig{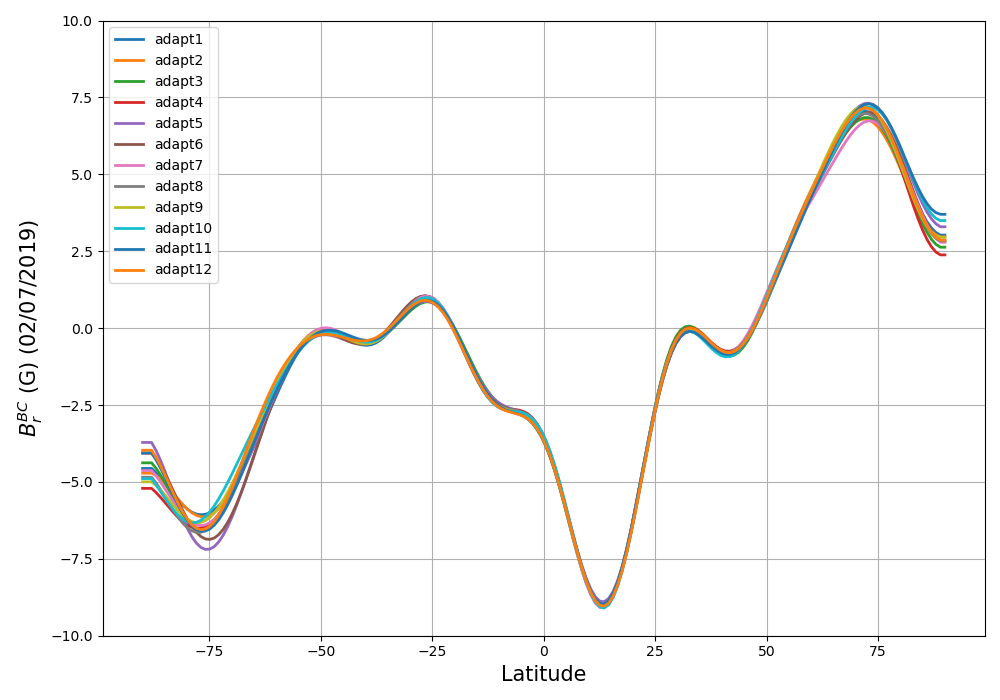}{0.49\textwidth}{(f) GONG-ADAPT pre-processed maps.}
    }
    \caption{Comparison of 1D cuts of the radial magnetic field $B_r$ at the longitude of the event ($2^{nd}$ of July 2019). On the left column (panels (a), (c), (e)), the cuts are made through the original magnetic maps. On the right column (panels (b), (d), (f)), the cuts are made through the pre-processed maps used as input for the simulations. The first row (panels (a) and (b)) shows the Carrington frame diachronic maps, the second row (panels (c) and (d)) shows the synchronic frame maps, the last row (panels (e) and (f)) the GONG-ADAPT realizations.}
    \label{fig:1d_cut_maps}
\end{figure}

The other point we want to stress is the question of the roles of the solar poles in space weather forecasts. This is important because most space weather models actually remove the solar poles, arguing that they are not relevant for forecasts at the Earth. However, it has been shown previously that the HCS location for example is very sensitive to the value of the polar field \citep{svalgaard1978}, and it is an important feature for space weather forecasts due to its possible interaction with CMEs \citep{Lavraud2014}. It is undeniable that saving precious computational time can help, however it is essential to quantify the impact of this decision. It may be justified for heliospheric propagators, since the polar boundary condition has little impact on the structures at Earth, but it is way more difficult to be sure for coronal models. That is why we want to focus specifically on this point in our study.

% 1D cut
We have shown in figure \ref{fig:std_bcfile} that the poles are actually the largest source of differences between all the various maps at the selected date. To show these differences in a more quantitative way in figure \ref{fig:1d_cut_maps}, we perform a 1D cut though all the maps at the Carrington longitude of the date we have chosen, which is around 315 degrees (panels (a), (c) and (e)). We also show the same cut after the pre-processing, to show what is actually used in the simulation (panels (b), (d) and (f)). The main difference is the amplitude of the magnetic field: before the pre-processing, the amplitudes range from -50 to 35 G, while afterwards they range from -9 to 7.5 G. The resolution is also affected as the pre-processing cuts off the smallest spatial structures. The first row (panels (a) and (b)) shows the Carrington frame diachronic maps, the second row (panels (c) and (d)) shows the synchronic frame maps, the last row (panels (e) and (f)) the GONG-ADAPT realizations. It is already visible from the original maps that the poles exhibit significant difference, but it is even more dominant after the pre-processing. It is then clear that the maps with which we obtain the best results are the ones that gather more magnetic field at the poles: the GONG-ADAPT maps (panel (f)) because of their flux-transport model, and the HMI maps (orange line in panel (b) and blue line in panel (d)), probably thanks to their high resolution. The GONG zero-corrected products also show some decent magnetic field at the poles (red line in panels (b) and (d)), which probably explains their good scores as well. Bad scores can also be related to bad assessment of polarity at the poles: both WSO and GONG mrmqs (blue and green lines in panel (b)) have extremely inaccurate extrapolations of the poles, with GONG mrmqs even having the wrong polarity at the southern pole, which explains why they get the worse scores. Too much magnetic field at the poles in combination with the numerical diffusion of our model may however lead to underestimating the equatorial regions, as we have seen that the GONG-ADAPT runs completely miss the equatorial coronal holes in a typical operational set-up (see figures \ref{fig:euv_ch} and \ref{fig:ch_comp_all}, and table \ref{tab:metrics}).

We have shown in section \ref{subsec:min_hcs} that depending on the input map, our simulations exhibited different shifts of the HCS. Since we have also shown in section \ref{subsec:input_br} that the biggest source of difference between the input maps was the treatment of the solar poles, we can assume that it is an important factor to explain this shift. It is also expected from \cite{svalgaard1978} that the magnetic field at the solar poles is going to impact the HCS, causing a shift of several degrees that can completely change its location at 1 AU with respect to Earth and hence change the geo-effectiveness and intensity of space weather events. Most of the differences between the maps we selected at minimum of activity also came exclusively from the poles, and this had very visible effects on the organization of the corona. In particular, the flux accumulation for the GONG-ADAPT map seems to cause our model in this standard operational set-up to miss the equatorial coronal holes contrary to other input maps, which are sources of high-speed streams that hit the Earth and trigger space weather events. This reinforces the importance of the ongoing mission Solar Orbiter, which will be the first imager to capture a global vision of the solar poles, hence helping the filling and calibration of the maps more accurately. With a combination of resolution and accuracy, we can combine the two advantages of HMI and GONG-ADAPT, and thus produce the best map that will be able to yield reliable simulations for forecasts.

\section{Conclusion}
\label{sec:conclusion}

We have tested the impact of the choice of the input magnetic map on the results of our coronal solar wind simulations using our new MHD implicit code COCONUT. To this end, we have selected a strategic date ($2^{nd}$ of July 2019) at minimum of activity in order to focus on the influence of the solar poles. This choice is recent enough for having a well-documented case and is during a total solar eclipse on Earth, allowing for having precise observations of the coronal structures at that time. We have gathered all 20 publicly available magnetic maps for this date from 4 different providers (WSO, HMI, GONG and GONG-ADAPT), spanning various resolutions and pole-filling techniques. We have pre-processed all maps the same way, with a spherical harmonics cut-off at $\ell_{max}=15$, which would be a standard pre-processing in space weather forecasting operational mode. In order to assess the quality of the resulting simulations, we have used three validation techniques with three different remote-sensing observations: we have estimated the magnetic field configuration (especially the shape and size of the streamers) from white-light total solar eclipse images, the open magnetic field lines repartition from EUV maps from SDO/AIA and the position of the HCS using white-light images from SoHO/LASCO/C2. We have also computed automatic metrics in order to evaluate automatically the quality of these comparisons.

What we have seen is that our model performs decently when using input from most maps, and allow for a comfortable visual comparison. However, we have obtained quite different results depending on the choice of the map, which shows that even at minimum of activity (i.e. event for quiet configurations) the input data has a strong impact. The quality of estimation for the streamers varies from 24\% to 85\%, with an average quality of about 60\%. Coronal holes estimation varies from 24\% to 88\% for the polar coronal holes (with an average of 80\% for the northern coronal hole, and 40\% for the southern coronal hole), and from 0\% to 65\% for the equatorial ones as some simulations completely fail to reproduce them. The HCS deviation from the SB estimate ranges on average from 4 to 12 degrees. We have tried to use these results in order to provide guidelines for using our model for space weather applications, which could probably be extended to other models with similar approximations (ideal MHD and polytropic heating) and boundary conditions. We can already estimate that a similar deviation of the HCS would be observed at 0.1 AU, which means that the input boundary condition for heliospheric propagators would definitively be affected. We have also assembled a scoreboard of the performances of our model for each map, which shows that with our model we should not use GONG mrbqs maps as they yield poor results. Instead, a better alternative would be the zero-corrected products such as GONG mrzqs and GONG mrnqs. Runs with GONG-ADAPT products perform very well, except for the equatorial coronal holes which are not reproduced at all. This could be a major issue for the inclusion of SIRs in the prevision. In the end, the best runs are actually the ones based on the HMI products, which should then become standard inputs for our model when used in space weather frameworks. We have linked these differences to the difference of resolution but also of treatment of the solar poles, as the flux-transport model from GONG-ADAPT is probably responsible for not reproducing the equatorial coronal holes in this operational set-up. This shows that the solar poles are needed to model accurately the first 20 solar radii and thus cannot be neglected without loss of information. This also highlights the importance of the ongoing Solar Orbiter mission that will provide more images of the solar poles in order to hopefully unify all these magnetic field measurements.

Of course, this study is just the first step towards better quantifying the requirements for space weather forecasts. It has proven that our model COCONUT is robust enough to take as input a large variety of maps, and has allowed us to identify the best maps to use to initialize it and provide inputs for space weather previsions, but there is still the need to see if these results can be generalized. We have studied only one minimum of activity, more cases would be needed to reach a conclusion for all minima. Another interesting point is whether these results still hold for maximum of activity cases: we actually expect the results to potentially vary a lot, since it is not the poles anymore that are driving the simulations, but rather the active regions, so probably that resolution and saturation effects would become more important. It is also not clear if these results hold for other numerical codes, although the previous comparison we did with Wind-Predict would suggest that at least for polytropic models we should find similar results \citep{Perri2022}. We will of course keep improving our model to be able to include more physics: the next key-points are the improvement of the modeling for the coronal heating in order to be able to have a bimodal distribution of the solar wind, as well as a multi-fluid treatment to be able to include a realistic transition region up to the chromosphere. Both these treatments will help include structures such as SIRs, and thus enable in-situ comparisons through coupling with heliospheric propagators such as EUHFORIA. In the end, we hope to be able to prove that our new coronal model not only helps to improve space weather forecasts of the wind structures, but also the transients propagating through this description of the interplanetary medium.

\begin{acknowledgements}
% Persons
The authors would like to thank Nicolas Poirier for providing the white-light SMB maps, and Jasmina Magdaleni\'c for useful discussions. 
% Fundings
This work has been granted by the AFOSR basic research initiative project FA9550-18-1-0093. 
This project has also received funding from the European Union’s Horizon 2020 research and innovation program under grant agreement No.~870405 (EUHFORIA 2.0) and the ESA project "Heliospheric modelling techniques“ (Contract No. 4000133080/20/NL/CRS).
F.Z.\ is supported by a postdoctoral mandate from KU Leuven Internal Funds  (PDMT1/21/028).
These results were also obtained in the framework of the projects
C14/19/089  (C1 project Internal Funds KU Leuven), G.0D07.19N  (FWO-Vlaanderen), SIDC Data Exploitation (ESA Prodex-12), and Belspo projects BR/165/A2/CCSOM and B2/191/P1/SWiM.
% Computers
The resources and services used in this work were provided by the VSC (Flemish Supercomputer Centre), funded by the Research Foundation - Flanders (FWO) and the Flemish Government.
% WSO
Wilcox Solar Observatory data used in this study was obtained via the web site \url{http://wso.stanford.edu} courtesy of J.T.\ Hoeksema.
The Wilcox Solar Observatory is currently supported by NASA.
% GONG
Data were acquired by GONG instruments operated by NISP/NSO/AURA/NSF with contribution from NOAA.
% HMI
HMI data are courtesy of the Joint Science Operations Center (JSOC) Science Data Processing team at Stanford University.
% ADAPT
This work utilizes data produced collaboratively between AFRL/ADAPT and NSO/NISP. \\
% Links
Data used in this study was obtained from the following websites: \\
WSO: \url{http://wso.stanford.edu/synopticl.html} \\
GONG: \url{https://gong2.nso.edu/archive/patch.pl?menutype=z} \\
HMI: \url{http://jsoc.stanford.edu/HMI/LOS_Synoptic_charts.html} \\
GONG-ADAPT: \url{https://gong.nso.edu/adapt/maps/}
\end{acknowledgements}

\appendix

\section{Complementary metric plots}
\label{app:metrics}

In this appendix, we will present some complementary plots that are briefly mentioned throughout the paper. The reason they were not included in the main paper is because they were too voluminous, with each time 20 subplots for the 20 cases considered.

\subsection{Input radial magnetic field maps}

\begin{figure}
    \centering
    \gridline{\fig{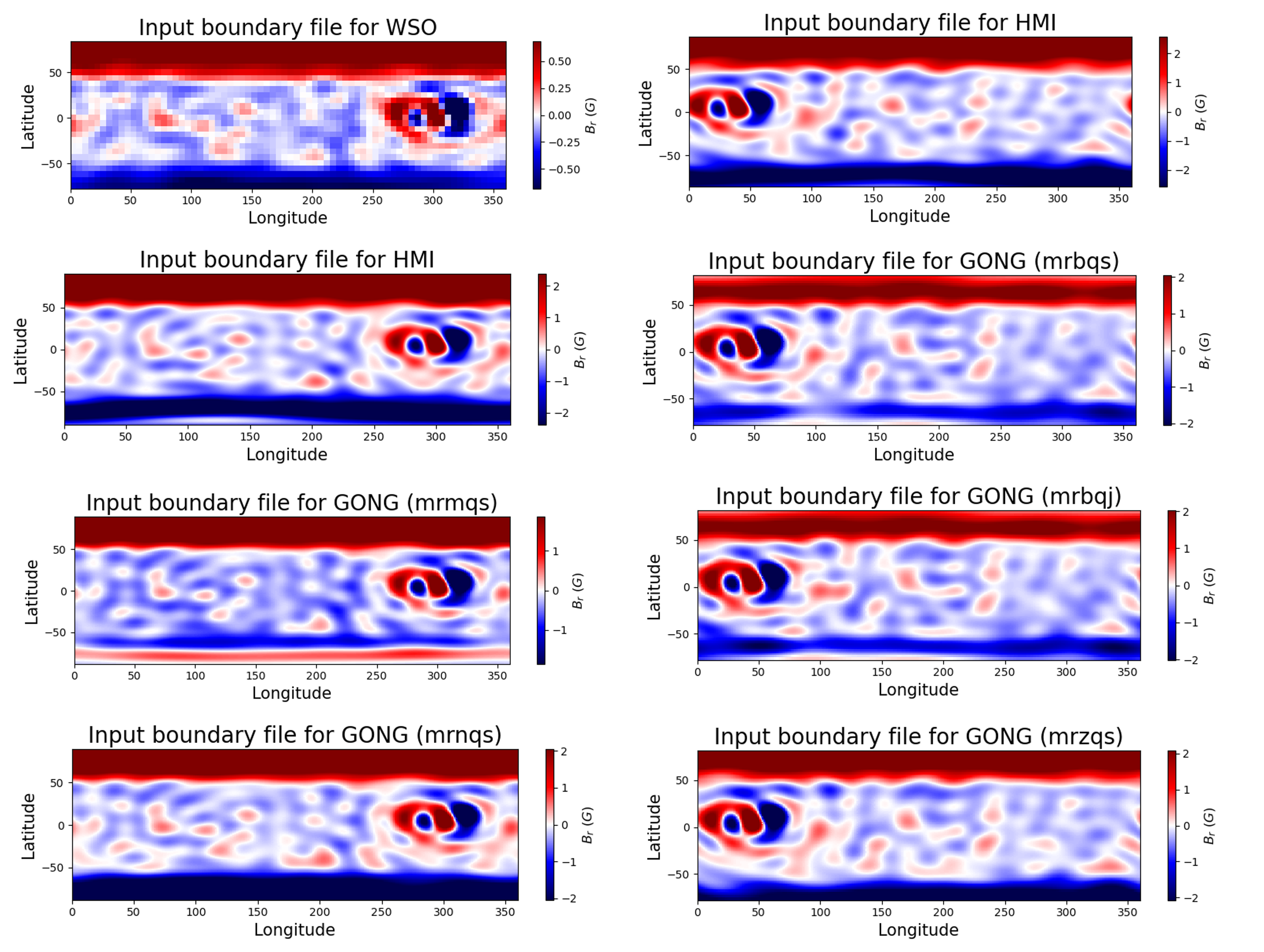}{0.8\textwidth}{}}
    \vspace{-1cm}
    \gridline{\fig{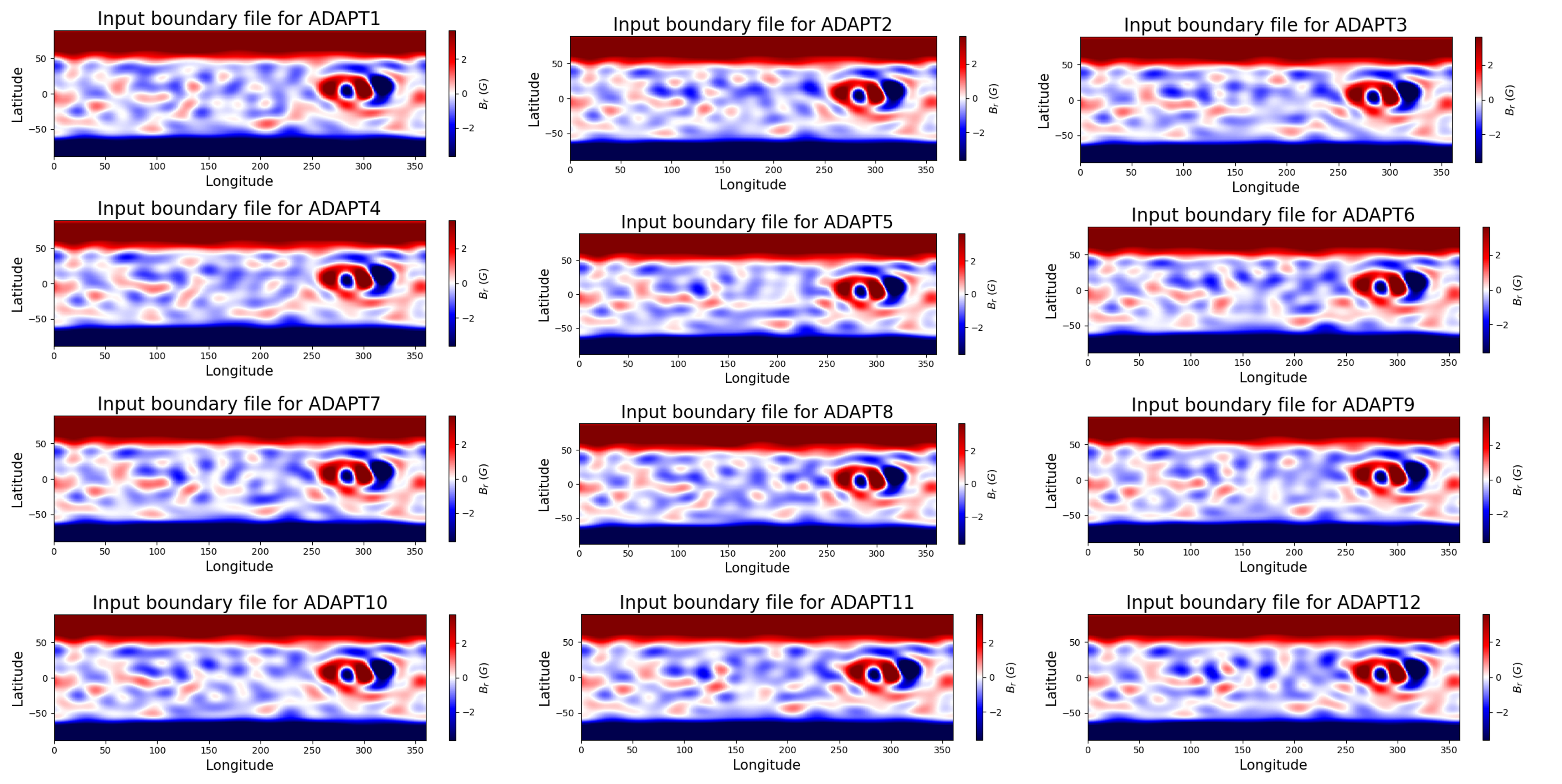}{\textwidth}{}}
    \caption{Radial magnetic field $B_r$ which is used as input boundary condition for the simulations after applying the map pre-processing. All maps shown here have been smoothened using a spherical harmonics decomposition with $\ell_{max}=15$. The column at the top-left shows maps in Carrington diachronic frame, the column at the top-right maps in synchronic frame, and the final block at the bottom shows all 12 realizations from the same GONG-ADAPT map. The color bar has been adjusted to show positive magnetic polarity in red, and negative polarity in blue. Each subplot has its own color bar to better show the differences in amplitude between the input fields.}
    \label{fig:bc_file_all}
\end{figure}

% Remind pre-processing
In section \ref{sec:mag_maps}, we have presented the various magnetic maps used for our simulations. However, as explained in section \ref{subsec:input_br}, we don't use directly the maps as they are, we apply a pre-processing step to them in order to use them as input in our code. We use a standard automatic pre-processing to simulate an operational framework without any optimization. To this end, we apply a spherical harmonics filtering with a cut-off at $\ell_{max}=15$. This results in the modified synoptic maps visible in figure \ref{fig:bc_file_all}. For each case, we show the resulting radial magnetic field $B_r$ map used directly as input to our simulations. We can see that the pre-processing smoothens the differences in resolution, with now only the WSO map showing a significant difference. This also reduces the difference in amplitude of the magnetic field: before, the maximum amplitude went up to 50 G with the HMI map, while now it reaches only 3 G for the GONG-ADAPT maps. This is because the pre-processing we chose cuts off the contribution of the small-scale structures that correspond to the active region. This allows us to focus more on the effects of the poles, as we wanted.

\subsection{Streamers overlap maps}

\begin{figure}
    \centering
    \gridline{\fig{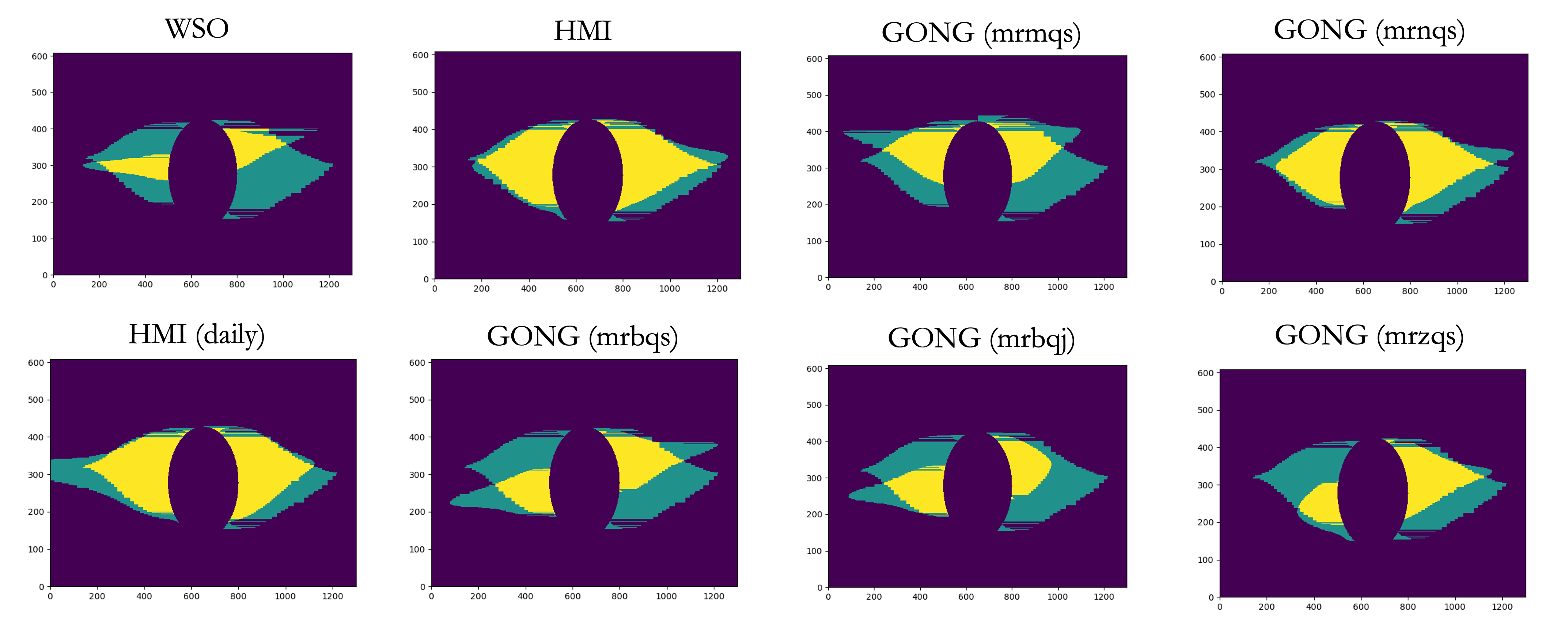}{\textwidth}{}}
    \vspace{-1cm}
    \gridline{\fig{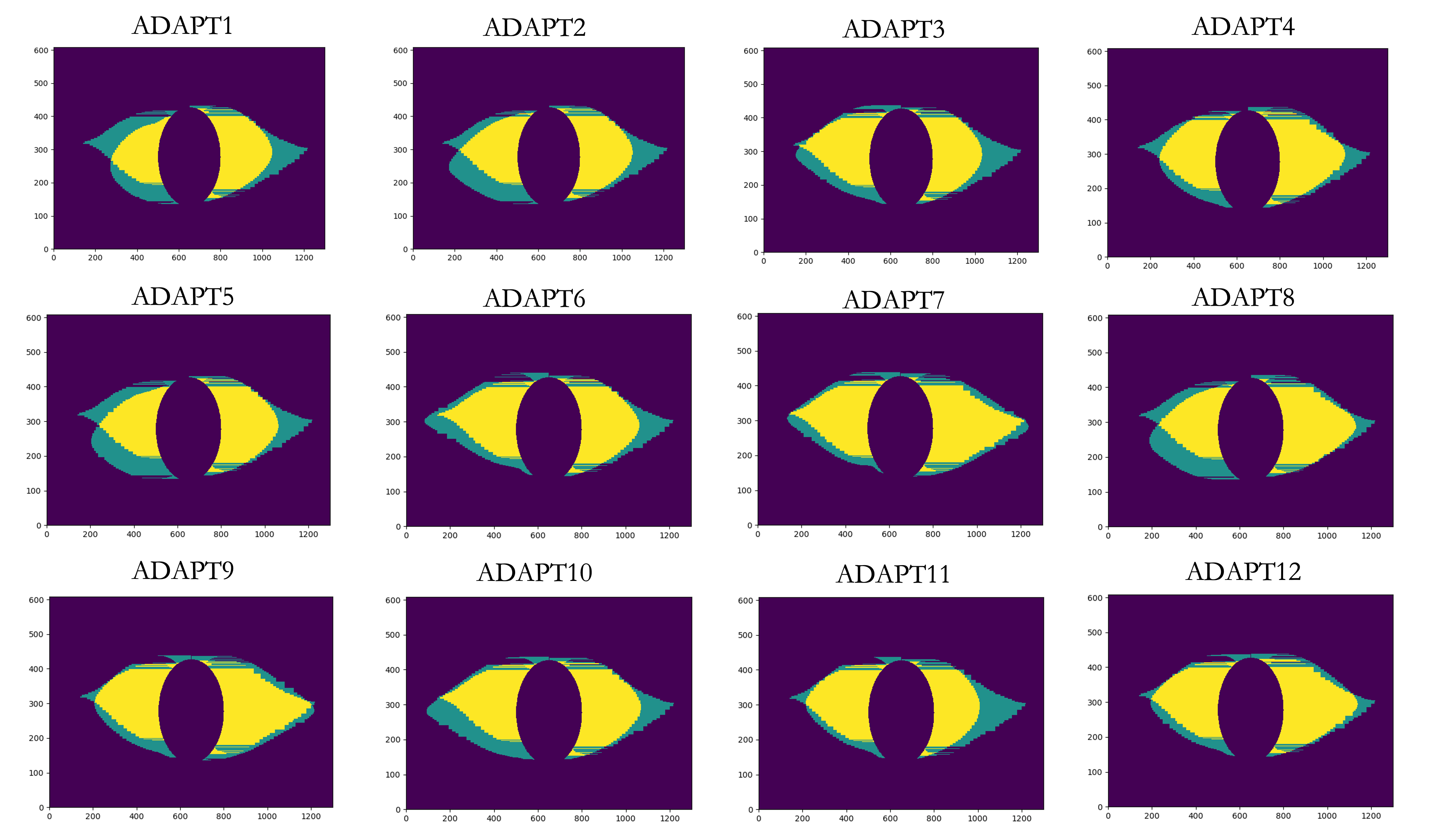}{\textwidth}{}}
    \caption{Maps of the streamers' coverage computation for each map simulation. An estimation of the streamer edges in the plane perpendicular to the observer line of sight is extracted manually from the white-light image of the eclipse and plotted with the streamer extracted from the simulation. Each pixel that does not belong within any of the streamers is in purple. Each pixel that belongs to one streamer is in green. Each pixel that belongs to the two streamers is in yellow.}
    \label{fig:streamers_comp_all}
\end{figure}

% Explain the method again
In section \ref{subsec:min_wl}, we have compared qualitatively the shape of the streamers we obtained from our MHD simulations to total solar eclipses white-light images that allow to estimate such structures. In section \ref{subsec:map_scores}, we have compiled a more quantitative score to be able to evaluate each map performance. We have explained the principle behind this metric, which is shown in figure \ref{fig:streamers_comp_all}. From the white-light image, we can extract a visual estimation of the shapes of the two streamers that are perpendicular to the line-of-sight. Then, we can compare it to the shape of the streamers extracted automatically from our numerical simulations by selecting the biggest closed magnetic field lines. We can then detect which pixel belongs within each streamer and obtain the maps shown in figure \ref{fig:streamers_comp_all}: if a pixel does not belong to any streamer, it is in purple; if it belongs to one streamer only, it is in green; if it belongs to both streamers (from the observations and from the simulations), it is in yellow. We then compare the percentage of yellow pixels with the number of pixels within the biggest streamer between observation and simulation. This allows us to avoid the case where the simulation streamers includes the observation streamer, which would yield 100\% coverage while some of the detection is false. If the observation streamer is bigger, then we compute the percentage of pixels detected. If the simulation streamer is bigger, then we compute the percentage of right detection over false detection. Of course, we would get a better estimate if there was a way to detect automatically the streamer's edge in the white-light picture, but the 3D projection to a 2D picture makes it still very challenging \citep{boe2020}. The next step would be to generate directly white-light emissions from the simulations, although this raises new problematics linked to the emission functions selected and the filters applied afterwards to see the structures \citep{Mikic2018}. Our procedure has the advantage of being semi-automatic and very universal since it relies directly on the magnetic field provided by the simulations.

\subsection{Coronal hole overlap maps}

\begin{figure}
    \centering
    \gridline{\fig{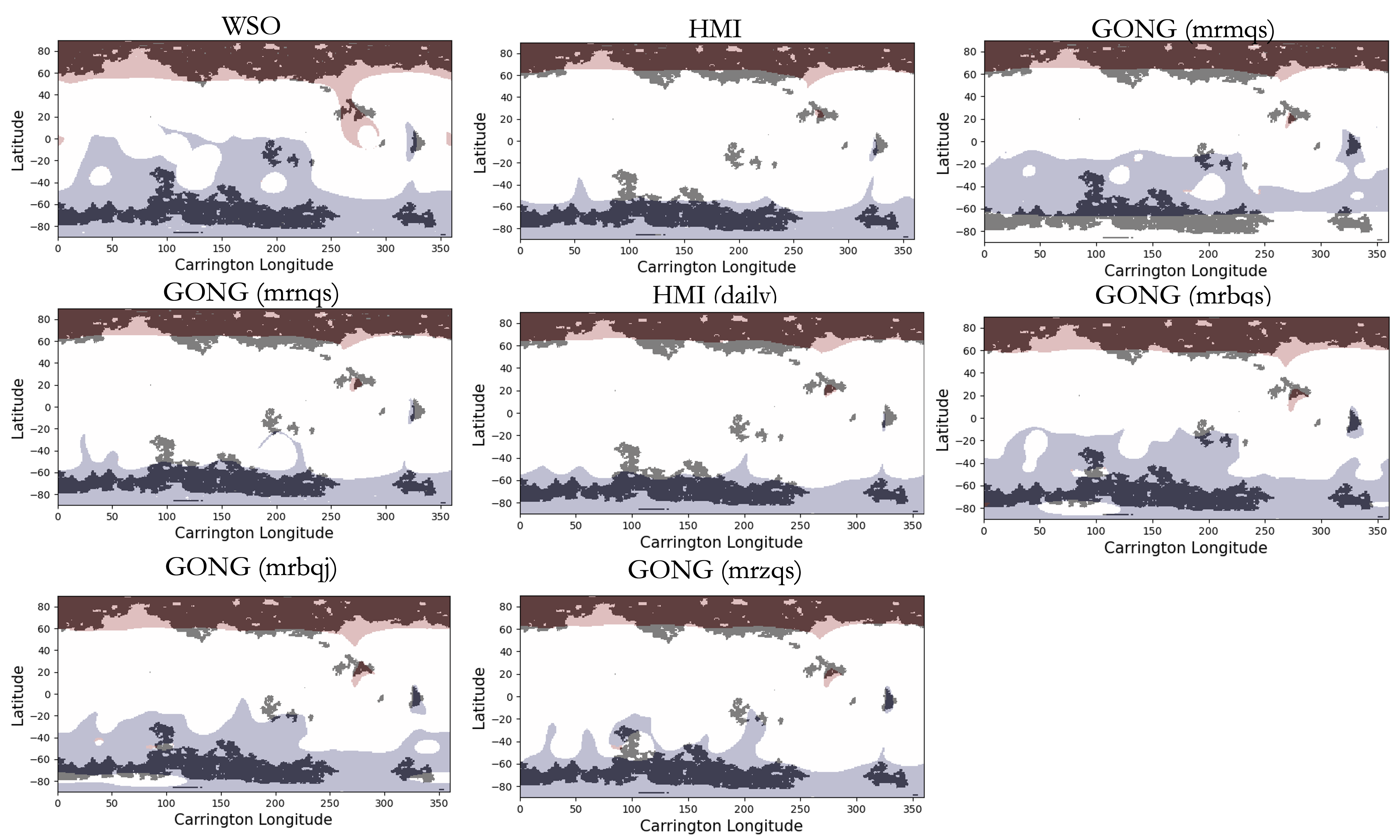}{0.8\textwidth}{}}
    \vspace{-1cm}
    \gridline{\fig{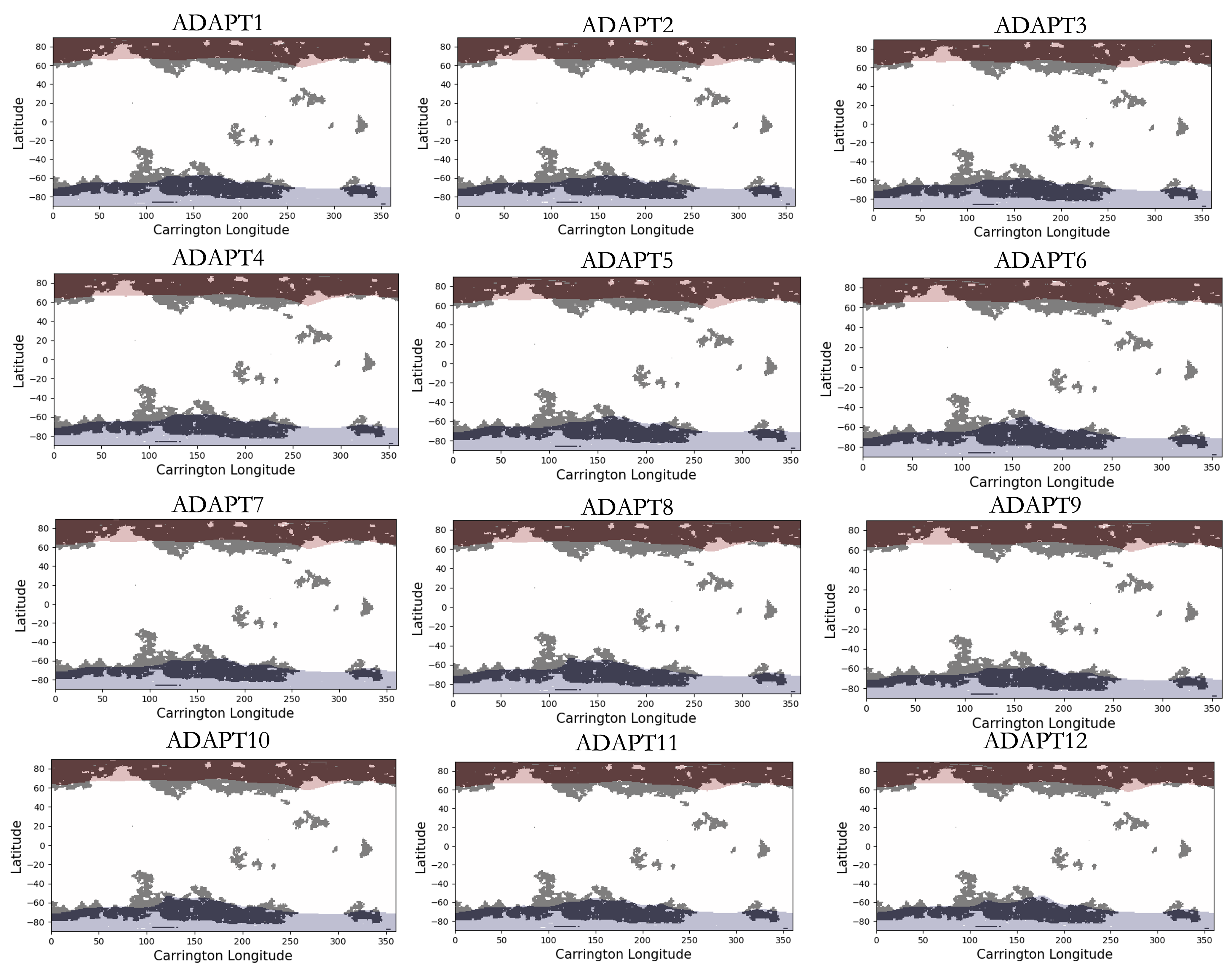}{0.8\textwidth}{}}
    \caption{Comparison between the coronal holes extracted from the EUV synoptic map and the MHD simulations for each map. The coronal hole contours are extracted using the EZSEG algorithm and plotted in gray. The coronal holes from the simulations are plotted in red (for positive polarity) and blue (for negative polarity).}
    \label{fig:ch_comp_all}
\end{figure}

% Explain EZSEG for coronal hole detection
% Explain coronal hole
% Explain comparison
In section \ref{subsec:min_ch}, we have compared visually the repartition of open magnetic field lines at the surface of the star with coronal hole maps derived from EUV emission from SDO/AIA (195 \AA channel). In section \ref{subsec:map_scores}, we once again need a more quantitative metric to be able to evaluate the performance of each map.
To do so, we first have to be able to extract automatically the pixels that belong to the coronal holes. We have done so by applying the EZSEG algorithm developed by Predictive Science Inc. \citep{Caplan2016ApJ}. The software is available as part of the EUV2CHM Matlab package from the Predictive Science Inc. website \footnote{\url{https://www.predsci.com/chd/}}. We have converted the algorithm in Python to be able to use it directly on our pipeline for the EUV synoptic map. This
algorithm uses an initial intensity threshold to acquire
coronal hole locations in an EUV image, and then
uses an area growing technique to define connected regions. This continues until a second intensity threshold is reached, or the condition for connectivity is not met. The dual-thresholds and connectivity conditions (essentially the number of consecutive pixels), are defined on input. We experimented with the optimal input parameters, and found that for this map the best result was obtained with a connectivity of 3 neighbors, a first threshold at 20 and a second threshold at 35. The resulting coronal hole detection can be seen in gray on all the panels in figure \ref{fig:ch_comp_all}. The coronal hole for the simulations were determined as said before by using seeds for the field lines and checking whether the field line would reach the outer boundary of the computational domain. With the simulation we can even link our coronal holes back to the polarity of the magnetic field. In figure \ref{fig:ch_comp_all}, we can then show the contours of our artificial coronal holes in red for associated positive polarity and in blue for negative associated polarity. We then computed the ratio of the number of pixels in both coronal hole detections to the number of pixels from the coronal holes from the simulation. That way, this percentage represents how accurate the coronal hole from the simulation is. We have separated polar and equatorial coronal holes by defining the equatorial region as being between -40 and 40 degrees in latitude. The quality of our comparison is limited by two factors. First, the EUV synoptic map has been reprojected from sinus latitudes to equally-spaced latitudes, which can generate some uncertainties at the poles. It seems that the southern pole in particular is badly affected, as we can see in figure \ref{fig:euv_ch}, which would explain why the southern coronal hole seems more disrupted and generates not so good scores. We are also limited by the pre-processing of the input map which removes some small-scale structures, but that way we are closer to operational results.

\bibliography{cf_mag}{}

\begin{thebibliography}{}
\expandafter\ifx\csname natexlab\endcsname\relax\def\natexlab#1{#1}\fi
\providecommand{\url}[1]{\href{#1}{#1}}

\bibitem[{{Alonso Asensio} {et~al.}(2019){Alonso Asensio}, {Alvarez Laguna},
  {Aissa}, {Poedts}, {Ozak}, \& {Lani}}]{Asensio2019}
{Alonso Asensio}, I., {Alvarez Laguna}, A., {Aissa}, M.~H., {et~al.} 2019,
  Computer Physics Communications, 239, 16

\bibitem[{Alvarez~Laguna {et~al.}(2016)Alvarez~Laguna, Lani, Deconinck,
  Mansour, \& Poedts}]{Laguna2016}
Alvarez~Laguna, A., Lani, A., Deconinck, H., Mansour, N., \& Poedts, S. 2016,
  Journal of Computational Physics, 318, doi:10.1016/j.jcp.2016.04.058

\bibitem[{{Alvarez Laguna} {et~al.}(2019){Alvarez Laguna}, {Ozak}, {Lani},
  {Mansour}, {Deconinck}, \& {Poedts}}]{Laguna2018}
{Alvarez Laguna}, A., {Ozak}, N., {Lani}, A., {et~al.} 2019, J. of Phys:
  Conference Series, 1031, 12th International Conference on Numerical Modeling
  of Space Plasma Flows, ASTRONUM 2017

\bibitem[{{Aschwanden}(2004)}]{Aschwanden2004}
{Aschwanden}, M.~J. 2004, {Physics of the Solar Corona. An Introduction}

\bibitem[{{Badman} {et~al.}(2022){Badman}, {Brooks}, {Poirier}, {Warren},
  {Petrie}, {Rouillard}, {Arge}, {Bale}, {de Pablos Aguero}, {Harra}, {Jones},
  {Kouloumvakos}, {Riley}, {Panasenco}, {Velli}, \& {Wallace}}]{Badman2022}
{Badman}, S.~T., {Brooks}, D.~H., {Poirier}, N., {et~al.} 2022, arXiv e-prints,
  arXiv:2201.11818

\bibitem[{Balan {et~al.}(2014)Balan, Skoug, Tulasi~Ram, Rajesh, Shiokawa,
  Otsuka, Batista, Ebihara, \& Nakamura}]{Balan2014cme}
Balan, N., Skoug, R., Tulasi~Ram, S., {et~al.} 2014, Journal of Geophysical
  Research: Space Physics, 119, 10

\bibitem[{Balay {et~al.}(1997)Balay, Gropp, McInnes, \&
  Smith}]{petsc-efficient}
Balay, S., Gropp, W.~D., McInnes, L.~C., \& Smith, B.~F. 1997, in Modern
  Software Tools in Scientific Computing, ed. E.~Arge, A.~M. Bruaset, \& H.~P.
  Langtangen (Birkh{\"{a}}user Press), 163--202

\bibitem[{Balay {et~al.}(2015{\natexlab{a}})Balay, Abhyankar, Adams, Brown,
  Brune, Buschelman, Dalcin, Eijkhout, Gropp, Kaushik, Knepley, McInnes, Rupp,
  Smith, Zampini, \& Zhang}]{petsc-web-page}
Balay, S., Abhyankar, S., Adams, M.~F., {et~al.} 2015{\natexlab{a}}, {PETS}c
  {W}eb page, \url{http://www.mcs.anl.gov/petsc}, , .
\newblock \url{http://www.mcs.anl.gov/petsc}

\bibitem[{Balay {et~al.}(2015{\natexlab{b}})Balay, Abhyankar, Adams, Brown,
  Brune, Buschelman, Dalcin, Eijkhout, Gropp, Kaushik, Knepley, McInnes, Rupp,
  Smith, Zampini, \& Zhang}]{petsc-user-ref}
---. 2015{\natexlab{b}}, {PETS}c Users Manual, Tech. Rep. ANL-95/11 - Revision
  3.6, Argonne National Laboratory.
\newblock \url{http://www.mcs.anl.gov/petsc}

\bibitem[{{Boe} {et~al.}(2020){Boe}, {Habbal}, \& {Druckm{\"u}ller}}]{boe2020}
{Boe}, B., {Habbal}, S., \& {Druckm{\"u}ller}, M. 2020, \apj, 895, 123

\bibitem[{{Bothmer} \& {Daglis}(2007)}]{Bothmer2007}
{Bothmer}, V., \& {Daglis}, I.~A. 2007, {Space Weather -- Physics and Effects},
  doi:10.1007/978-3-540-34578-7

\bibitem[{{Brchnelova} {et~al.}(2022{\natexlab{a}}){Brchnelova}, {Perri},
  {Brchnelova}, {Baratishvili}, {Kuzma}, {Zhang}, {Lani}, \&
  {Poedts}}]{Brchnelova2022b}
{Brchnelova}, M., {Perri}, B., {Brchnelova}, M., {et~al.} 2022{\natexlab{a}},
  Journal of Plasma Physics, to be submitted

\bibitem[{{Brchnelova} {et~al.}(2022{\natexlab{b}}){Brchnelova}, {Zhang},
  {Leitner}, {Perri}, {Lani}, \& {Poedts}}]{Brchnelova2022a}
{Brchnelova}, M., {Zhang}, F., {Leitner}, P., {et~al.} 2022{\natexlab{b}},
  Journal of Plasma Physics, 88, 905880205

\bibitem[{{Brun} \& {Browning}(2017)}]{Brun2017}
{Brun}, A.~S., \& {Browning}, M.~K. 2017, Living Reviews in Solar Physics, 14,
  4

\bibitem[{{Caplan} {et~al.}(2016){Caplan}, {Downs}, \&
  {Linker}}]{Caplan2016ApJ}
{Caplan}, R.~M., {Downs}, C., \& {Linker}, J.~A. 2016, \apj, 823, 53

\bibitem[{{Caplan} {et~al.}(2021){Caplan}, {Downs}, {Linker}, \&
  {Mikic}}]{Caplan2021}
{Caplan}, R.~M., {Downs}, C., {Linker}, J.~A., \& {Mikic}, Z. 2021, \apj, 915,
  44

\bibitem[{{Chhiber} {et~al.}(2021){Chhiber}, {Usmanov}, {Matthaeus}, \&
  {Goldstein}}]{Chhiber2021}
{Chhiber}, R., {Usmanov}, A.~V., {Matthaeus}, W.~H., \& {Goldstein}, M.~L.
  2021, arXiv e-prints, arXiv:2107.11657

\bibitem[{{Chorin}(1997)}]{chorin1997}
{Chorin}, A.~J. 1997, Journal of Computational Physics, 135, 118

\bibitem[{{Cranmer}(2009)}]{Cranmer2009}
{Cranmer}, S.~R. 2009, Living Reviews in Solar Physics, 6, 3

\bibitem[{{Cranmer} \& {Winebarger}(2019)}]{Cranmer2019}
{Cranmer}, S.~R., \& {Winebarger}, A.~R. 2019, \araa, 57, 157

\bibitem[{Dedner {et~al.}(2002)Dedner, Kemm, Kröner, Munz, Schnitzer, \&
  Wesenberg}]{Dedner2002}
Dedner, A., Kemm, F., Kröner, D., {et~al.} 2002, Journal of Computational
  Physics, 175, 645.
\newblock
  \url{https://www.sciencedirect.com/science/article/pii/S002199910196961X}

\bibitem[{{Fargette} {et~al.}(2021){Fargette}, {Lavraud}, {Rouillard},
  {R{\'e}ville}, {Dudok De Wit}, {Froment}, {Halekas}, {Phan}, {Malaspina},
  {Bale}, {Kasper}, {Louarn}, {Case}, {Korreck}, {Larson}, {Pulupa}, {Stevens},
  {Whittlesey}, \& {Berthomier}}]{Fargette2021}
{Fargette}, N., {Lavraud}, B., {Rouillard}, A.~P., {et~al.} 2021, \apj, 919, 96

\bibitem[{{Gopalswamy} {et~al.}(2012){Gopalswamy}, {Yashiro}, {M{\"a}kel{\"a}},
  {Michalek}, {Shibasaki}, \& {Hathaway}}]{Gopalswamy2012}
{Gopalswamy}, N., {Yashiro}, S., {M{\"a}kel{\"a}}, P., {et~al.} 2012, \apjl,
  750, L42

\bibitem[{{Hickmann} {et~al.}(2015){Hickmann}, {Godinez}, {Henney}, \&
  {Arge}}]{Hickmann2015}
{Hickmann}, K.~S., {Godinez}, H.~C., {Henney}, C.~J., \& {Arge}, C.~N. 2015,
  \solphys, 290, 1105

\bibitem[{{Ito} {et~al.}(2010){Ito}, {Tsuneta}, {Shiota}, {Tokumaru}, \&
  {Fujiki}}]{Ito2010}
{Ito}, H., {Tsuneta}, S., {Shiota}, D., {Tokumaru}, M., \& {Fujiki}, K. 2010,
  \apj, 719, 131

\bibitem[{{Jian} {et~al.}(2015){Jian}, {MacNeice}, {Taktakishvili}, {Odstrcil},
  {Jackson}, {Yu}, {Riley}, {Sokolov}, \& {Evans}}]{Jian2015}
{Jian}, L.~K., {MacNeice}, P.~J., {Taktakishvili}, A., {et~al.} 2015, Space
  Weather, 13, 316

\bibitem[{{Jin} {et~al.}(2022){Jin}, {Nitta}, \& {Cohen}}]{meng2022}
{Jin}, M., {Nitta}, N.~V., \& {Cohen}, C. M.~S. 2022, arXiv e-prints,
  arXiv:2202.07214

\bibitem[{Kimpe {et~al.}(2005)Kimpe, Lani, Quintino, Poedts, \&
  Vandewalle}]{Kimpe2005}
Kimpe, D., Lani, A., Quintino, T., Poedts, S., \& Vandewalle, S. 2005, in
  Recent Advances in Parallel Virtual Machine and Message Passing Interface,
  ed. B.~Di~Martino, D.~Kranzlm{\"u}ller, \& J.~Dongarra (Berlin, Heidelberg:
  Springer Berlin Heidelberg), 520--527

\bibitem[{Lani {et~al.}(2005)Lani, Quintino, Kimpe, Deconinck, Vandewalle, \&
  Poedts}]{Lani2005}
Lani, A., Quintino, T., Kimpe, D., {et~al.} 2005, in Computational Science --
  ICCS 2005, ed. V.~S. Sunderam, G.~D. van Albada, P.~M.~A. Sloot, \& J.~J.
  Dongarra (Berlin, Heidelberg: Springer Berlin Heidelberg), 279--286

\bibitem[{Lani {et~al.}(2006)Lani, Quintino, Kimpe, Deconinck, Vandewalle, \&
  Poedts}]{Lani2006}
Lani, A., Quintino, T., Kimpe, D., {et~al.} 2006, Scientific Programming, 14,
  doi:10.1155/2006/393058

\bibitem[{Lani {et~al.}(2013)Lani, Villedieu, Bensassi, Kapa, Vymazal, Yalim,
  \& Panesi}]{Lani2013}
Lani, A., Villedieu, N., Bensassi, K., {et~al.} 2013, in AIAA 2013-2589, 21th
  AIAA CFD Conference, San Diego (CA)

\bibitem[{{Lani} {et~al.}(2014){Lani}, {Yalim}, \& S.}]{LaniGPU}
{Lani}, A., {Yalim}, M.~S., \& S., P. 2014, Computer Physics Communications,
  185, 2538

\bibitem[{{Lavraud} \& {Rouillard}(2014)}]{Lavraud2014}
{Lavraud}, B., \& {Rouillard}, A. 2014, in Nature of Prominences and their Role
  in Space Weather, ed. B.~{Schmieder}, J.-M. {Malherbe}, \& S.~T. {Wu}, Vol.
  300, 273--284

\bibitem[{{Li} {et~al.}(2021){Li}, {Feng}, \& {Wei}}]{Li2021}
{Li}, H., {Feng}, X., \& {Wei}, F. 2021, Journal of Geophysical Research (Space
  Physics), 126, e28870

\bibitem[{Linker {et~al.}(2017)Linker, Caplan, Downs, Riley, Miki{\'c},
  Lionello, Henney, Arge, Liu, Derosa, Yeates, \& Owens}]{Linker2017}
Linker, J.~A., Caplan, R.~M., Downs, C., {et~al.} 2017, The Astrophysical
  Journal, 848, 70.
\newblock \url{https://doi.org/10.3847/1538-4357/aa8a70}

\bibitem[{{Lionello} {et~al.}(2009){Lionello}, {Linker}, \&
  {Miki{\'c}}}]{Lionello2009}
{Lionello}, R., {Linker}, J.~A., \& {Miki{\'c}}, Z. 2009, \apj, 690, 902

\bibitem[{Maneva {et~al.}(2017)Maneva, Laguna, Lani, \& Poedts}]{Maneva2017}
Maneva, Y.~G., Laguna, A.~A., Lani, A., \& Poedts, S. 2017, The Astrophysical
  Journal, 836, 197.
\newblock \url{https://doi.org/10.3847/1538-4357/aa5b83}

\bibitem[{{McComas} {et~al.}(2008){McComas}, {Ebert}, {Elliott}, {Goldstein},
  {Gosling}, {Schwadron}, \& {Skoug}}]{McComas2008}
{McComas}, D.~J., {Ebert}, R.~W., {Elliott}, H.~A., {et~al.} 2008, \grl, 35,
  L18103

\bibitem[{{McComas} {et~al.}(2003){McComas}, {Elliott}, {Schwadron}, {Gosling},
  {Skoug}, \& {Goldstein}}]{McComas2003}
{McComas}, D.~J., {Elliott}, H.~A., {Schwadron}, N.~A., {et~al.} 2003, \grl,
  30, 1517

\bibitem[{{Miki{\'c}} {et~al.}(2018){Miki{\'c}}, {}, {Downs}, {Linker},
  {Caplan}, {Mackay}, {Upton}, {Riley}, {Lionello}, {T{\"o}r{\"o}k}, {Titov},
  {Wijaya}, {Druckm{\"u}ller}, {Pasachoff}, \& {Carlos}}]{Mikic2018}
{Miki{\'c}}, {}, Z., {Downs}, C., {et~al.} 2018, Nature Astronomy, 2, 913

\bibitem[{{Miki{\'c}} {et~al.}(1999){Miki{\'c}}, {Linker}, {Schnack},
  {Lionello}, \& {Tarditi}}]{Mikic1999}
{Miki{\'c}}, Z., {Linker}, J.~A., {Schnack}, D.~D., {Lionello}, R., \&
  {Tarditi}, A. 1999, Physics of Plasmas, 6, 2217

\bibitem[{{Odstrcil}(2003)}]{Odstrcil2003}
{Odstrcil}, D. 2003, Advances in Space Research, 32, 497

\bibitem[{{Owens} \& {Forsyth}(2013)}]{Owens2013}
{Owens}, M.~J., \& {Forsyth}, R.~J. 2013, Living Reviews in Solar Physics, 10,
  5

\bibitem[{{Owens} {et~al.}(2008){Owens}, {Spence}, {McGregor}, {Hughes},
  {Quinn}, {Arge}, {Riley}, {Linker}, \& {Odstrcil}}]{Owens2008}
{Owens}, M.~J., {Spence}, H.~E., {McGregor}, S., {et~al.} 2008, Space Weather,
  6, S08001

\bibitem[{{Parenti} {et~al.}(2022){Parenti}, {R{\'e}ville}, {Brun}, {Pinto},
  {Auch{\`e}re}, {Buchlin}, {Perri}, \& {Strugarek}}]{Parenti2022}
{Parenti}, S., {R{\'e}ville}, V., {Brun}, A.~S., {et~al.} 2022, \apj, 929, 75

\bibitem[{{Perri} {et~al.}(2018){Perri}, {Brun}, {R{\'e}ville}, \&
  {Strugarek}}]{Perri2018}
{Perri}, B., {Brun}, A.~S., {R{\'e}ville}, V., \& {Strugarek}, A. 2018, Journal
  of Plasma Physics, 84, 765840501

\bibitem[{{Perri} {et~al.}(2022){Perri}, {Leitner}, {Brchnelova},
  {Baratishvili}, {Kuzma}, {Zhang}, {Lani}, \& {Poedts}}]{Perri2022}
{Perri}, B., {Leitner}, P., {Brchnelova}, M., {et~al.} 2022, \apj, submitted

\bibitem[{{Petrie}(2013)}]{Petrie2013}
{Petrie}, G.~J.~D. 2013, \apj, 768, 162

\bibitem[{{Petrie}(2015)}]{Petrie2015}
{Petrie}, G. J.~D. 2015, Living Reviews in Solar Physics, 12, 5

\bibitem[{{Pinto} \& {Rouillard}(2017)}]{Pinto2017}
{Pinto}, R.~F., \& {Rouillard}, A.~P. 2017, \apj, 838, 89

\bibitem[{{Pirjola}(2005)}]{Pirjola2005}
{Pirjola}, R. 2005, Advances in Space Research, 36, 2231

\bibitem[{{Poedts} {et~al.}(2020){Poedts}, {Kochanov}, {Lani}, {Scolini},
  {Verbeke}, {Hosteaux}, {Chan{\'e}}, {Deconinck}, {Mihalache}, {Diet},
  {Heynderickx}, {De Keyser}, {De Donder}, {Crosby}, {Echim}, {Rodriguez},
  {Vansintjan}, {Verstringe}, {Mampaey}, {Horne}, {Glauert}, {Jiggens}, {Keil},
  {Glover}, {Deprez}, \& {Luntama}}]{Poedts2020_vswmc}
{Poedts}, S., {Kochanov}, A., {Lani}, A., {et~al.} 2020, Journal of Space
  Weather and Space Climate, 10, 14

\bibitem[{{Poedts, Stefaan} {et~al.}(2020){Poedts, Stefaan}, {Lani, Andrea},
  {Scolini, Camilla}, {Verbeke, Christine}, {Wijsen, Nicolas}, {Lapenta,
  Giovanni}, {Laperre, Brecht}, {Millas, Dimitrios}, {Innocenti, Maria Elena},
  {Chan\'e, Emmanuel}, {Baratashvili, Tinatin}, {Samara, Evangelia}, {Van der
  Linden, Ronald}, {Rodriguez, Luciano}, {Vanlommel, Petra}, {Vainio, Rami},
  {Afanasiev, Alexandr}, {Kilpua, Emilia}, {Pomoell, Jens}, {Sarkar, Ranadeep},
  {Aran, Angels}, {Sanahuja, Blai}, {Paredes, Josep M.}, {Clarke, Ellen},
  {Thomson, Alan}, {Rouilard, Alexis}, {Pinto, Rui F.}, {Marchaudon,
  Aur\'elie}, {Blelly, Pierre-Louis}, {Gorce, Blandine}, {Plotnikov, Illya},
  {Kouloumvakos, Athanasis}, {Heber, Bernd}, {Herbst, Konstantin}, {Kochanov,
  Andrey}, {Raeder, Joachim}, \& {Depauw, Jan}}]{Poedts2020_euhforia}
{Poedts, Stefaan}, {Lani, Andrea}, {Scolini, Camilla}, {et~al.} 2020, J. Space
  Weather Space Clim., 10, 57.
\newblock \url{https://doi.org/10.1051/swsc/2020055}

\bibitem[{{Poirier} {et~al.}(2021){Poirier}, {Rouillard}, {Kouloumvakos},
  {Przybylak}, {Fargette}, {Pobeda}, {R{\'e}ville}, {Pinto}, {Indurain}, \&
  {Alexandre}}]{Poirier2021}
{Poirier}, N., {Rouillard}, A.~P., {Kouloumvakos}, A., {et~al.} 2021, Frontiers
  in Astronomy and Space Sciences, 8, 84

\bibitem[{{Pomoell} \& {Poedts}(2018)}]{Pomoell2018}
{Pomoell}, J., \& {Poedts}, S. 2018, Journal of Space Weather and Space
  Climate, 8, A35

\bibitem[{{Pulkkinen}(2007)}]{Pulkkinen2007}
{Pulkkinen}, T. 2007, Living Reviews in Solar Physics, 4, 1

\bibitem[{Reames(2013)}]{Reames_2013}
Reames, D.~V. 2013, Space Science Reviews, 175, 53.
\newblock \url{https://doi.org/10.1007%2Fs11214-013-9958-9}

\bibitem[{{Reames}(2021)}]{Reames2021}
{Reames}, D.~V. 2021, {Solar Energetic Particles. A Modern Primer on
  Understanding Sources, Acceleration and Propagation}, Vol. 978,
  doi:10.1007/978-3-030-66402-2

\bibitem[{{R{\'e}ville} {et~al.}(2015){R{\'e}ville}, {Brun}, {Strugarek},
  {Matt}, {Bouvier}, {Folsom}, \& {Petit}}]{Reville2015b}
{R{\'e}ville}, V., {Brun}, A.~S., {Strugarek}, A., {et~al.} 2015, \apj, 814, 99

\bibitem[{{R{\'e}ville} {et~al.}(2020){R{\'e}ville}, {Velli}, {Panasenco},
  {Tenerani}, {Shi}, {Badman}, {Bale}, {Kasper}, {Stevens}, {Korreck},
  {Bonnell}, {Case}, {de Wit}, {Goetz}, {Harvey}, {Larson}, {Livi},
  {Malaspina}, {MacDowall}, {Pulupa}, \& {Whittlesey}}]{Reville2020}
{R{\'e}ville}, V., {Velli}, M., {Panasenco}, O., {et~al.} 2020, \apjs, 246, 24

\bibitem[{{Riley} \& {Ben-Nun}(2021)}]{Riley2021}
{Riley}, P., \& {Ben-Nun}, M. 2021, Space Weather, 19, e02775

\bibitem[{{Riley} {et~al.}(2019){Riley}, {Linker}, {Mikic}, {Caplan}, {Downs},
  \& {Thumm}}]{Riley2019}
{Riley}, P., {Linker}, J.~A., {Mikic}, Z., {et~al.} 2019, \apj, 884, 18

\bibitem[{{Riley} {et~al.}(2014){Riley}, {Ben-Nun}, {Linker}, {Mikic},
  {Svalgaard}, {Harvey}, {Bertello}, {Hoeksema}, {Liu}, \&
  {Ulrich}}]{riley2014}
{Riley}, P., {Ben-Nun}, M., {Linker}, J.~A., {et~al.} 2014, \solphys, 289, 769

\bibitem[{Saad \& Schultz(1986)}]{Saad1986}
Saad, Y., \& Schultz, M.~H. 1986, SIAM Journal on Scientific and Statistical
  Computing, 7, 856

\bibitem[{{Samara} {et~al.}(2022){Samara}, {Laperre}, {Kieokaew}, {Temmer},
  {Verbeke}, {Rodriguez}, {Magdaleni{\'c}}, \& {Poedts}}]{Samara2022}
{Samara}, E., {Laperre}, B., {Kieokaew}, R., {et~al.} 2022, \apj, 927, 187

\bibitem[{{Samara} {et~al.}(2021){Samara}, {Pinto}, {Magdaleni{\'c}}, {Wijsen},
  {Jer{\v{c}}i{\'c}}, {Scolini}, {Jebaraj}, {Rodriguez}, \&
  {Poedts}}]{Samara2021}
{Samara}, E., {Pinto}, R.~F., {Magdaleni{\'c}}, J., {et~al.} 2021, \aap, 648,
  A35

\bibitem[{{Schrijver} {et~al.}(2015){Schrijver}, {Kauristie}, {Aylward},
  {Denardini}, {Gibson}, {Glover}, {Gopalswamy}, {Grande}, {Hapgood},
  {Heynderickx}, {Jakowski}, {Kalegaev}, {Lapenta}, {Linker}, {Liu},
  {Mandrini}, {Mann}, {Nagatsuma}, {Nandy}, {Obara}, {Paul O'Brien}, {Onsager},
  {Opgenoorth}, {Terkildsen}, {Valladares}, \& {Vilmer}}]{Schrijver2015}
{Schrijver}, C.~J., {Kauristie}, K., {Aylward}, A.~D., {et~al.} 2015, Advances
  in Space Research, 55, 2745

\bibitem[{{Shen} {et~al.}(2022){Shen}, {Shen}, {Xu}, {Liu}, {Feng}, \&
  {Wang}}]{Shen2022}
{Shen}, F., {Shen}, C., {Xu}, M., {et~al.} 2022, Reviews of Modern Plasma
  Physics, 6, 8

\bibitem[{{Shiota} {et~al.}(2014){Shiota}, {Kataoka}, {Miyoshi}, {Hara}, {Tao},
  {Masunaga}, {Futaana}, \& {Terada}}]{Shiota2014}
{Shiota}, D., {Kataoka}, R., {Miyoshi}, Y., {et~al.} 2014, Space Weather, 12,
  187

\bibitem[{{Singh} {et~al.}(2018){Singh}, {Yalim}, \& {Pogorelov}}]{Singh2018}
{Singh}, T., {Yalim}, M.~S., \& {Pogorelov}, N.~V. 2018, \apj, 864, 18

\bibitem[{Svalgaard(2006)}]{svalgaard2006good}
Svalgaard, L. 2006, in Presentation at 2006 SHINE Workshop

\bibitem[{{Svalgaard} \& {Wilcox}(1978)}]{svalgaard1978}
{Svalgaard}, L., \& {Wilcox}, J.~M. 1978, \araa, 16, 429

\bibitem[{{Temmer}(2021)}]{Temmer2021}
{Temmer}, M. 2021, Living Reviews in Solar Physics, 18, 4

\bibitem[{{T{\'o}th} {et~al.}(2012){T{\'o}th}, {van der Holst}, {Sokolov}, {De
  Zeeuw}, {Gombosi}, {Fang}, {Manchester}, {Meng}, {Najib}, {Powell}, {Stout},
  {Glocer}, {Ma}, \& {Opher}}]{Toth2012}
{T{\'o}th}, G., {van der Holst}, B., {Sokolov}, I.~V., {et~al.} 2012, Journal
  of Computational Physics, 231, 870

\bibitem[{{Tsuneta} {et~al.}(2008){Tsuneta}, {Ichimoto}, {Katsukawa}, {Lites},
  {Matsuzaki}, {Nagata}, {Orozco Su{\'a}rez}, {Shimizu}, {Shimojo}, {Shine},
  {Suematsu}, {Suzuki}, {Tarbell}, \& {Title}}]{Tsuneta2008}
{Tsuneta}, S., {Ichimoto}, K., {Katsukawa}, Y., {et~al.} 2008, \apj, 688, 1374

\bibitem[{{Ulrich}(1992)}]{Ulrich1992}
{Ulrich}, R.~K. 1992, in Astronomical Society of the Pacific Conference Series,
  Vol.~26, Cool Stars, Stellar Systems, and the Sun, ed. M.~S. {Giampapa} \&
  J.~A. {Bookbinder}, 265

\bibitem[{{van der Holst} {et~al.}(2014){van der Holst}, {Sokolov}, {Meng},
  {Jin}, {Manchester}, {T{\'o}th}, \& {Gombosi}}]{vanderHolst2014}
{van der Holst}, B., {Sokolov}, I.~V., {Meng}, X., {et~al.} 2014, \apj, 782, 81

\bibitem[{{Vidotto} {et~al.}(2018){Vidotto}, {Lehmann}, {Jardine}, \&
  {Pevtsov}}]{Vidotto2018}
{Vidotto}, A.~A., {Lehmann}, L.~T., {Jardine}, M., \& {Pevtsov}, A.~A. 2018,
  \mnras, 480, 477

\bibitem[{{Virtanen} \& {Mursula}(2017)}]{virtanen2017}
{Virtanen}, I., \& {Mursula}, K. 2017, \aap, 604, A7

\bibitem[{{Wagner} {et~al.}(2022){Wagner}, {Asvestari}, {Temmer}, {Heinemann},
  \& {Pomoell}}]{Wagner2022}
{Wagner}, A., {Asvestari}, E., {Temmer}, M., {Heinemann}, S.~G., \& {Pomoell},
  J. 2022, \aap, 657, A117

\bibitem[{{Wallace} {et~al.}(2019){Wallace}, {Arge}, {Pattichis},
  {Hock-Mysliwiec}, \& {Henney}}]{Wallace2019}
{Wallace}, S., {Arge}, C.~N., {Pattichis}, M., {Hock-Mysliwiec}, R.~A., \&
  {Henney}, C.~J. 2019, \solphys, 294, 19

\bibitem[{{Wang} {et~al.}(2007){Wang}, {Sheeley}, \& {Rich}}]{Wang2007}
{Wang}, Y.~M., {Sheeley}, N.~R., J., \& {Rich}, N.~B. 2007, \apj, 658, 1340

\bibitem[{{Wu} {et~al.}(2006){Wu}, {Wang}, {Liu}, \& {Hoeksema}}]{Wu2006}
{Wu}, S.~T., {Wang}, A.~H., {Liu}, Y., \& {Hoeksema}, J.~T. 2006, \apj, 652,
  800

\bibitem[{{Yalim} {et~al.}(2017){Yalim}, {Pogorelov}, {Singh}, \&
  {Liu}}]{Yalim2017}
{Yalim}, M.~S., {Pogorelov}, N., {Singh}, T., \& {Liu}, Y. 2017, in AGU Fall
  Meeting Abstracts, Vol. 2017, SH23D--2698

\bibitem[{{Yalim} {et~al.}(2011){Yalim}, {Vanden Abeele}, {Lani}, {Quintino},
  \& {Deconinck}}]{Yalim}
{Yalim}, M.~S., {Vanden Abeele}, D., {Lani}, A., {Quintino}, T., \&
  {Deconinck}, H. 2011, Journal of Computational Physics, 230, 6136

\bibitem[{{Yeates} {et~al.}(2018){Yeates}, {Amari}, {Contopoulos}, {Feng},
  {Mackay}, {Miki{\'c}}, {Wiegelmann}, {Hutton}, {Lowder}, {Morgan}, {Petrie},
  {Rachmeler}, {Upton}, {Canou}, {Chopin}, {Downs}, {Druckm{\"u}ller},
  {Linker}, {Seaton}, \& {T{\"o}r{\"o}k}}]{Yeates2018}
{Yeates}, A.~R., {Amari}, T., {Contopoulos}, I., {et~al.} 2018, \ssr, 214, 99

\end{thebibliography}
\bibliographystyle{aasjournal}

\end{document}